\documentclass[aps,pre,showpacs,showkeys,onecolumn,superscriptaddress]{revtex4}
\usepackage[dvips]{graphicx}
\usepackage[dvips]{color}
\usepackage{hyperref,breakurl,amsmath,amssymb,pst-node,eepic}

\newcommand{\ie}{\emph{i.e.}, }
\newcommand{\ii}{\mathrm{i}}
\newcommand{\ee}{\mathrm{e}}
\newcommand{\dd}{\mathrm{d}}
\newcommand{\DD}{\mathrm{D}}
\newcommand{\re}{\mathrm{Re}}
\newcommand{\im}{\mathrm{Im}}
\newcommand{\Tr}{\mathrm{Tr}}

\newcommand{\Id}{\mathbf{1}}
\newcommand{\Zero}{\mathbf{0}}

\newcommand{\la}{\left\langle}
\newcommand{\ra}{\right\rangle}
\newcommand{\tot}{\textrm{tot.}}

\begin{document}

\title{Eigenvalues and Singular Values\\of Products of
Rectangular Gaussian Random Matrices}

\author{Z. Burda}
\email{zdzislaw.burda@uj.edu.pl}
\affiliation{Marian Smoluchowski Institute of Physics,
Jagiellonian University, Reymonta 4, 30--059 Krak\'{o}w, Poland}
\affiliation{Mark Kac Complex Systems Research Centre,
Jagiellonian University, Reymonta 4, 30--059 Krak\'{o}w, Poland}
\author{A. Jarosz}
\email{jedrekjarosz@gmail.com}
\affiliation{The Henryk Niewodnicza\'{n}ski Institute of Nuclear Physics, Polish Academy of Sciences, Radzikowskiego 152, 31--342 Krak\'{o}w, Poland}
\author{G. Livan}
\email{giacomo.livan@pv.infn.it}
\affiliation{Dipartimento di Fisica Nucleare e Teorica,
Universit\`{a} degli Studi di Pavia, Via Bassi 6, 27100 Pavia,
Italy}
\affiliation{Istituto Nazionale di Fisica Nucleare, Sezione di
Pavia, Via Bassi 6, 27100 Pavia, Italy}
\author{M. A. Nowak}
\email{nowak@th.if.uj.edu.pl}
\affiliation{Marian Smoluchowski Institute of Physics,
Jagiellonian University, Reymonta 4, 30--059 Krak\'{o}w, Poland}
\affiliation{Mark Kac Complex Systems Research Centre,
Jagiellonian University, Reymonta 4, 30--059 Krak\'{o}w, Poland}
\author{A. Swiech}
\email{artur.swiech@uj.edu.pl}
\affiliation{Marian Smoluchowski Institute of Physics,
Jagiellonian University, Reymonta 4, 30--059 Krak\'{o}w, Poland}

\date{\today}

\setlength{\parindent}{4ex}
\setlength{\parskip}{1.5ex plus 0.5ex minus 0.2ex}
\topmargin=0cm

\begin{abstract}
We derive exact analytic expressions for the distributions of
eigenvalues
and singular values for the product of an arbitrary number of
independent rectangular Gaussian random matrices in the limit of
large matrix dimensions. We show that
they both have power--law behavior
at zero and determine the corresponding powers. We also propose
a heuristic form of finite size corrections to these expressions
which very well approximates
the distributions for matrices of finite dimensions.
\end{abstract}


\maketitle


\section{Introduction}
\label{s:Introduction}

Spectral analysis of the products of random matrices is a powerful tool in several domains of statistical physics, allowing, for example, to study Lyapunov exponents for disordered and chaotic dynamical systems~\cite{VULPIANI}. It is also useful in a class of problems related to multiplicative matrix--valued noncommutative diffusion processes~\cite{LAUTRUP}. Several applications go beyond physics, as for instance, those related to the stability analysis of ecological systems~\cite{ECO} or to telecommunication applications based on the scattering of electromagnetic waves on random obstacles~\cite{LIGHT,tv}. In many of those cases, some exact analytic results were obtained for relatively small matrices. Interestingly, quite often analytic calculations are possible under another limit --- the limit of matrix dimensions tending to infinity. Examples include products of pseudounitary matrices, representing transfer matrices in mesoscopic wires~\cite{BENAKKER}, large $N$ Wilson loops in Yang--Mills theory~\cite{NN,JW,BN} or multiplicative diffusion of infinitely large complex and/or Hermitian matrices~\cite{gjjn,LNW}. In most of these cases, the reason why the exact spectral distribution is within the reach of analytic methods is due to a link to free random variable calculus~\cite{s,vdn}, which is a very powerful technique. This is also why the spectra of products of large random matrices represent a challenge for mathematicians~\cite{bbcc,bg}. In this paper, we generalize the analysis of the product of large, square, random Gaussian matrices, performed in~\cite{bjw}, to the product of rectangular matrices. In particular, we study the product
\begin{equation}\label{eq:ProductDefinition}
\mathbf{P} \equiv \mathbf{A}_{1} \mathbf{A}_{2} \ldots
\mathbf{A}_{L}
\end{equation}
of $L \geq 1$ independent, rectangular, large, random Gaussian
matrices \smash{$\mathbf{A}_{l}$}, $l = 1 , 2 , \ldots , L$,
of dimensions $N_{l} \times N_{l+1}$. We are interested in the
eigenvalue and singular value density of $\mathbf{P}$ in
the limit $N_{L+1} \rightarrow \infty$ and
\begin{equation}\label{eq:ThermodynamicLimit}
R_{l} \equiv \frac{N_{l}}{N_{L+1}} = \textrm{finite} , \qquad
\textrm{for} \qquad l = 1 , 2 , \ldots , {L+1}.
\end{equation}
In other words, all matrix dimensions grow to infinity at fixed
rates and, obviously, \smash{$R_{L+1} = 1$}. The product
$\mathbf{P}$ is
a matrix of dimensions $N_1 \times N_{L+1}$ and has eigenvalues
only
if it is a square matrix: $N_1 = N_{L+1}$.


We assume the matrices \smash{$\mathbf{A}_{l}$} in
the product (\ref{eq:ProductDefinition}) to be complex Gaussian
matrices drawn randomly from the ensemble defined by
the probability measure
\begin{equation}\label{eq:RectangularGGMeasure}
\dd \mu \left( \mathbf{A}_{l} \right) \propto \ee^{-
\frac{\sqrt{N_{l} N_{l + 1}}}{\sigma_{l}^{2}} \Tr \left(
\mathbf{A}_{l}^{\dagger} \mathbf{A}_{l} \right)} \DD
\mathbf{A}_{l},
\end{equation}
where
\smash{$\DD \mathbf{A} \equiv \prod_{a,b} \dd ( \re [ \mathbf{A}
]_{a b} ) \dd ( \im [ \mathbf{A} ]_{a b} )$} is a flat measure. 
A normalization constant, fixed by the condition $\int \dd \mu
\left( \mathbf{A} \right) =1$, is omitted.
This is the simplest generalization of the Girko--Ginibre
ensemble \cite{g1,g2,g3} to
rectangular matrices. The $\sigma_l$ parameters set the scale
for the
Gaussian fluctuations in \smash{$\mathbf{A}_{l}$}'s. The entries
of each
matrix \smash{$\mathbf{A}_{l}$} can be viewed as independent
centered Gaussian
random variables, the variance of the real and imaginary parts
being proportional
to $\sigma_l^2$ and inversely proportional to the square root of
the
number $N_l N_{l+1}$  of elements in the matrix.


The eigenvalue density of the product (\ref{eq:ProductDefinition}) of square Gaussian matrices was calculated in~\cite{bjw} while the singular value distribution was determined in~\cite{bbcc, bg,MULLER}. The eigenvalue density was derived using a planar diagrammatic method for non--Hermitian matrices~\cite{jnpwz,jnpz1,jnpz2,gjjn}, while the singular value density was obtained using Free Random Variables calculus~\cite{ns,v,vdn}. Both techniques work in the infinite matrix size limit. After explaining notation~(Section II) and listing the main results of the paper~(Section III), we shall follow those same methods to derive the corresponding results for the product of rectangular matrices. In~Section IV, we present a diagrammatic derivation of the moment generating function for the product $\bf{P}$. In~Section~V, using the tools of Free Random Variables calculus, we obtain the moment generating function for $\bf{Q}={\bf P}^{\dagger}{\bf P}$, recovering results given in~\cite{MULLER}. Section~VI concludes the paper with a discussion on particular applications of our results and possible generalizations.


\section{Generalities}


Let us spend a few words on the notations to be used in this
paper. The eigenvalue density $\rho_{\mathbf{X}}(\lambda)$ of a
Hermitian matrix $\mathbf{X}$ is a real function of real
argument, while in the case of a non--Hermitian matrix it is a
real function of complex argument. In the latter case we shall
write
$\rho_{\mathbf{X}}(\lambda,\bar{\lambda})$ and treat $\lambda$
and its conjugate $\bar{\lambda}$ as independent variables.

In the Hermitian case, the eigenvalue density can be
computed from a Green's function $G_{\mathbf{X}} ( z )$~\cite{m,gmw} which contains
the same information as the density itself:
\begin{equation}\label{eq:MeanSpectralDensityFromGreensFunction}\rho_{\mathbf{X}} ( \lambda ) = - \frac{1}{\pi} \lim_{\epsilon
\to 0^{+}} \im G_{\mathbf{X}} ( \lambda + \ii \epsilon ) .
\end{equation}

For a non--Hermitian matrix, the corresponding Green's function
$G_\mathbf{X}(z,\bar{z})$ is non--holomorphic and therefore
we shall write it explicitly as a function of $z$ and $\bar{z}$.
In this
case the eigenvalue distribution is reconstructed from the
Green's function
as~\cite{scss,fs,fks}
\begin{equation}\label{eq:MeanSpectralDensityFromNonHolomorphicGreensFunction}
\rho_{\mathbf{X}} ( z , \overline{z} ) = \frac{1}{\pi}
\frac{\partial}{\partial \overline{z}} G_{\mathbf{X}} ( z ,
\overline{z} ).
\end{equation}
Actually, this equation reduces to
(\ref{eq:MeanSpectralDensityFromGreensFunction}) when the
non--holomorphic region
shrinks to a cut along the real axis, as it happens for
Hermitian matrices.
The Green's function $G_{\mathbf{X}} (z)$ for a Hermitian matrix
is written as a function of a single argument since everywhere
except on the cut one has
$\partial_{\bar{z}} G_{\mathbf{X}} = 0$, and thus it is
$\bar{z}$--independent.


In many applications it is often convenient to use the moment
generating function, or $M$--transform, which is closely related
to the Green's function: $M_{\mathbf{X}}(z) = z
G_{\mathbf{X}}(z) - 1$. For a Hermitian matrix ${\mathbf{X}}$ one has
\begin{equation} \label{Mtrans}
M_{\mathbf{X}} ( z ) = \sum_{n \geq 1} \frac{m_n}{z^{n}} =
\sum_{n \geq 1} \frac{1}{z^{n}} \int \rho_{\mathbf{X}} (\lambda)
\lambda^n \dd \lambda,
\end{equation}
where the $m_n$'s are the moments of the eigenvalue density. If the matrix
${\mathbf{X}}$ is of finite dimensions $N \times N$,
the moments are given by $m_n =\frac{1}{N} \langle \Tr \mathbf{X}^n \rangle$.
The moment generating function encodes the same information as
the Green's function $G_{\mathbf{X}}(z) =
z^{-1}M_{\mathbf{X}}(z) + z^{-1}$. Thus, one can calculate the
corresponding eigenvalue distribution from $M_{\mathbf{X}}(z)$.


One can also introduce a similar function for non--Hermitian
matrices: $M_{\mathbf{X}}(z,\bar{z}) = z
G_{\mathbf{X}}(z,\bar{z}) - 1$.
In this case, however, it does not play the role of a moment
generating function anymore, since now one also has
mixed moments $\left\langle \Tr \mathbf{X}^n (\mathbf{X}^\dagger)^k
\right\rangle$, which in general depend on the ordering of
$\mathbf{X}$ and $\mathbf{X}^\dagger$ in the product under the trace.

The situation is slightly simplified when the $M$--transform is a
spherically
symmetric function: $M_{\mathbf{X}}(z,\bar{z}) ={\cal
M}_\mathbf{X}(|z|^2)$. In this case equation
(\ref{eq:MeanSpectralDensityFromNonHolomorphicGreensFunction})
can be cast into the form
\begin{equation} \label{rM}
\rho_{\mathbf{X}} ( z, \bar{z} ) = \frac{1}{\pi} {\cal
M}'_\mathbf{X}(|z|^2) +
f \delta^2(z,\bar{z})
\end{equation}
where ${\cal M}'_\mathbf{X}$ is the first derivative of ${\cal
M}_\mathbf{X}$
and $f=1+{\cal M}_\mathbf{X}(0)$ is a constant representing the
fraction of zero modes. In this case, the eigenvalue
distribution is spherically symmetric as well (see for example
\cite{bjw} for the product of square matrices). As we shall see
later, this is also going to be the case for the product
(\ref{eq:ProductDefinition})
of rectangular Gaussian matrices
(\ref{eq:RectangularGGMeasure}).


\section{Results}
\label{s:Results}


The matrix $\mathbf{P}$ (\ref{eq:ProductDefinition}) has
eigenvalues only if it is square, while it has singular
values for any rectangular shape.
As a matter of fact, its singular values can be determined as the
square roots of the non--zero eigenvalues of the matrix
\begin{equation}\label{eq:SingularValuesDefinition}
\mathbf{Q} \equiv \mathbf{P}^{\dagger} \mathbf{P}
\end{equation}
or, alternatively, of the matrix $\mathbf{R} = \mathbf{P}
\mathbf{P}^{\dagger}$.
$\mathbf{Q}$ and $\mathbf{R}$ are Hermitian, and they have
non--negative spectra
which differ only in the zero modes.


The main finding of this paper is that the eigenvalue
distribution and the $M$--transform of the product
(\ref{eq:ProductDefinition})
are spherically symmetric. We shall show the $M$--transform to
satisfy the $L$--th order polynomial equation
\begin{equation}\label{eq:MPBasicEquation}
\prod_{l = 1}^{L} \left( \frac{{\cal M}_{\mathbf{P}}
(|z|^2)}{R_{l}} + 1 \right) = \frac{| z |^{2}}{\sigma^{2}},
\end{equation}
where the scale parameter is $\sigma=\sigma_1 \sigma_2 \ldots
\sigma_L$. When all of the matrices involved are square, this
equation reproduces the results in \cite{bjw}.


An analogous equation for $\mathbf{Q}$ reads
\begin{equation}\label{eq:MQBasicEquation}
\sqrt{R_{1}} \frac{M_{\mathbf{Q}} ( z ) + 1}{M_{\mathbf{Q}} ( z
)} \prod_{l = 1}^{L} \left( \frac{M_{\mathbf{Q}} ( z )}{R_{l}} +
1 \right) = \frac{z}{\sigma^{2}} .
\end{equation}
It is an algebraic equation of order $( L + 1 )$, and it was first obtained in~\cite{MULLER} in the context of wireless telecommunication. Equations (\ref{eq:MPBasicEquation}) and (\ref{eq:MQBasicEquation}) are strikingly similar. They actually differ only by the prefactor in front of the product. Moreover, the free argument in the first equation is $|z|^2$, while $z$
in the second one. This observation represents the second main result of this paper. Since conventions used in telecommunication theory and in physics differ a bit, in Section~V we rederive equation (\ref{eq:MQBasicEquation}) for completeness.

When $\mathbf{P}$ is a square matrix, then $R_1=1$ and the square root at the
beginning of equation (\ref{eq:MQBasicEquation}) can be omitted.
When the product of square matrices is considered, all of the
$R_l$'s become equal to unity and the two equations take the
following form:
\begin{equation} \label{Requal1}
\left({\cal M}_{\mathbf{P}} (|z|^2) + 1 \right)^L = \frac{| z
|^{2}}{\sigma^{2}} \quad , \qquad
M_{\mathbf{Q}}^{-1} ( z ) \left(M_{\mathbf{Q}} ( z ) + 1
\right)^{L+1}
= \frac{z}{\sigma^{2}} \ .
\end{equation}
Equations \eqref{eq:MPBasicEquation} can be easily rewritten in
terms of the corresponding Green's
functions (see the previous section). If one does that and then
applies the prescriptions in
(\ref{eq:MeanSpectralDensityFromNonHolomorphicGreensFunction})
and
(\ref{eq:MeanSpectralDensityFromGreensFunction}) respectively,
it becomes clear that
\begin{equation}
\rho_{\mathbf{P}} ( \lambda , \overline{\lambda} ) \sim |
\lambda |^{-2 \frac{L-1}{L}} , \qquad
\mbox{and} \qquad
\rho_{\mathbf{Q}} ( \lambda ) \sim \lambda^{- \frac{L}{L + 1}} ,
\qquad \textrm{as} \qquad \lambda \to 0 .
\end{equation}
In the more general case of rectangular matrices, when solving
equations (\ref{eq:MPBasicEquation}) and
(\ref{eq:MQBasicEquation}) for the Green's functions, one can
then see that only
those brackets in which $R_l=1$ contribute to the singularity at
zero, while all others approach a constant for $z\rightarrow 0$.
Thus, the eigenvalue density displays the following singularity
\begin{equation}\label{eq:RhoPSingularityAtZero}
\rho_{\mathbf{P}} ( \lambda , \overline{\lambda} ) \sim |
\lambda |^{-2 \frac{s-1}{s}} , \qquad \textrm{as} \qquad \lambda
\to 0 ,
\end{equation}
where $s$ is the number of those ratios among $R_1,\ldots, R_L$
which are exactly equal to unity. On the other hand, the eigenvalue
density of $\mathbf{Q}$ behaves as
\begin{equation}\label{eq:RhoQSingularityAtZero}
\rho_{\mathbf{Q}} ( \lambda ) \sim \lambda^{- \frac{s}{s + 1}} ,
\qquad \textrm{as} \qquad \lambda \to 0 .
\end{equation}

The third result we want to mention here
is a heuristic form for the finite size corrections to the
eigenvalue distribution.
For a large but finite order of magnitude $N$ of the matrices
involved, the eigenvalue distribution is still
spherically symmetric. So, let $\rho_N( r )$ denote the radial
profile of
this distribution, where $r= | \lambda |$.
As we shall show, the evolution of the radial shape with the
size $N$
is very well described by a simple multiplicative correction:
\begin{equation}
\rho_N( r ) \equiv \rho( r ) \frac{1}{2} \mathrm{erfc} \left( q
( r - \sigma ) \sqrt{N} \right).
\end{equation}
In the $N\rightarrow \infty$ limit the correction becomes a step
function,
so that $\rho_\infty(r) = \rho(r) $ for $r\le \sigma$ and
$\rho_\infty(r)=0$
otherwise. The shape of the limiting radial distribution
$\rho(r)$ comes from the solution of
(\ref{eq:MPBasicEquation}). This type of finite size corrections
can be derived analytically for Girko--Ginibre matrices
\cite{fh,k,ks}.
Here we show that it also works very well for the eigenvalues of
the
product of Gaussian matrices. It is very generic and possibly
applies
to other random matrix ensembles with spherically symmetric
eigenvalue densities.


\section{The Eigenvalues of a Product of Rectangular Gaussian
Random Matrices}
\label{s:TheEigenvaluesOfAProductOfRectangularGaussianRandomMatrices}

In this section, we present a derivation of the main
result of this article, equation (\ref{eq:MPBasicEquation}), a
realization of it in the case $L=2$, and numerical simulations
to confirm our findings. To this end, we employ a technique for
summing planar diagrams, the Dyson--Schwinger equations (more
precisely described in \cite{gjjn,bjw}), extended to a non--Hermitian
framework.


The evaluation of the Green's function for a \emph{product}
$\mathbf{P}$ of random ensembles by means of planar
diagrammatics is non--linear w.r.t. the constituent matrices. It
is possible to linearize the problem by means of the following
trick~\cite{gjjn,bjw}. Consider the following block matrix:
\begin{equation}\label{eq:EIG01}
\widetilde{\mathbf{P}} \equiv \left( \begin{array}{ccccc} \Zero
& \mathbf{A}_{1} & \Zero & \ldots & \Zero \\ \Zero & \Zero &
\mathbf{A}_{2} & \ldots & \Zero \\ \vdots & \vdots & \vdots &
\ddots & \vdots \\ \Zero & \Zero & \Zero & \ldots &
\mathbf{A}_{L - 1} \\ \mathbf{A}_{L} & \Zero & \Zero & \ldots &
\Zero \end{array} \right).
\end{equation}
It is a matrix of dimensions \smash{$N_{\tot} \times
N_{\tot}$}, where \smash{$N_{\tot} \equiv N_{1} + N_{2} +
\ldots + N_{L}$}. It is important to notice that the $L$-th power of
$\widetilde{\mathbf{P}}$ is a block diagonal matrix~\cite{gjjn,bjw}
\begin{equation}
\widetilde{\mathbf{P}}^L = \left( \begin{array}{ccccc}
\mathbf{B}_{1} & \Zero & \Zero & \ldots & \Zero \\
\Zero & \mathbf{B}_{2} & \Zero & \ldots & \Zero \\
\Zero & \Zero & \mathbf{B}_{3} & \ldots & \Zero \\
\vdots & \vdots & \vdots & \ddots & \vdots \\
\Zero & \Zero & \Zero & \ldots & \mathbf{B}_L
\end{array} \right).
\end{equation}
with square blocks $\mathbf{B}_1 = \mathbf{A}_1 \mathbf{A}_2 \ldots
\mathbf{A}_{L-1} \mathbf{A}_L$,
$\mathbf{B}_2 =  \mathbf{A}_2 \ldots \mathbf{A}_L \mathbf{A}_1$, $\ldots \ $, being
cyclicly permuted products of $\mathbf{A}_1$,  $\mathbf{A}_2, \ldots, \mathbf{A}_L$. All these blocks have identical non--zero eigenvalues. They differ only
in the number of eigenvalues, which may vary from block to block. The first diagonal block $\mathbf{B}_1$ is equal to the product $\mathbf{P}$ (\ref{eq:ProductDefinition}). Taking into account that this block has dimensions $N_1 \times N_1$, while the total matrix $\widetilde{\mathbf{P}}^L$ has dimensions \smash{$N_{\tot} \times N_{\tot}$}, one can easily
deduce the following relation between the $M$--transforms of $\mathbf{P}$
and \smash{$\widetilde{\mathbf{P}}$}:
\begin{equation}\label{eq:EIG07a}
M_{\widetilde{\mathbf{P}}} ( w , \overline{w} ) = \frac{L
N_{1}}{N_{\tot}} M_{\mathbf{P}} \left( w^{L} , \overline{w}^{L}
\right) .
\end{equation}
The importance of this relation relies in the fact that one can use it to calculate \smash{$M_{\mathbf{P}}( z , \overline{z} )$} from \smash{$M_{\widetilde{\mathbf{P}}} ( w , \overline{w} )$}. The latter can be calculated using Dyson--Schwinger equations, since the matrix $\widetilde{\mathbf{P}}$ is linear w.r.t. the constituent matrices $\mathbf{A}_i$.


The first step in writing the Dyson--Schwinger equations is to
know the propagators of the random matrix in question, namely
\smash{$\widetilde{\mathbf{P}}$}, or more precisely its
``duplicated'' version:
\begin{equation}\label{eq:EIG09}
\widetilde{\mathbf{P}}^{\DD} = \left( \begin{array}{cc}
\widetilde{\mathbf{P}} & \Zero \\ \Zero &
\widetilde{\mathbf{P}}^{\dagger} \end{array} \right) .
\end{equation}
We shall think of it as a four--block matrix, each block being an
$L \times L$ block matrix. We shall denote the $L \times L$
block indices in these four blocks by $l m$ (upper left corner),
\smash{$l \overline{m}$} (upper right), \smash{$\overline{l} m$}
(lower left), \smash{$\overline{l} \overline{m}$} (lower right),
each one covering the range $1 , 2 , \ldots , L$; for example
\smash{$[ \widetilde{\mathbf{P}}^{\DD} ]_{\overline{2}
\overline{1}} = \mathbf{A}_{1}^{\dagger}$}. All the other
matrices involved shall inherit this same structure. For
instance,
\begin{equation}\label{eq:EIG10}
\mathbf{G}^{\DD} = \left( \begin{array}{c|c} \mathbf{G}^{w w} &
\mathbf{G}^{w \overline{w}} \\ \hline \mathbf{G}^{\overline{w}
w} & \mathbf{G}^{\overline{w} \overline{w}} \end{array} \right)
= \left( \begin{array}{cccc|cccc} \left[ \mathbf{G}^{\DD}
\right]_{1 1} & \left[ \mathbf{G}^{\DD} \right]_{1 2} & \ldots &
\left[ \mathbf{G}^{\DD} \right]_{1 L} & \left[ \mathbf{G}^{\DD}
\right]_{1 \overline{1}} & \left[ \mathbf{G}^{\DD} \right]_{1
\overline{2}} & \ldots & \left[ \mathbf{G}^{\DD} \right]_{1
\overline{L}} \\ \left[ \mathbf{G}^{\DD} \right]_{2 1} & \left[
\mathbf{G}^{\DD} \right]_{2 2} & \ldots & \left[
\mathbf{G}^{\DD} \right]_{2 L} & \left[ \mathbf{G}^{\DD}
\right]_{2 \overline{1}} & \left[ \mathbf{G}^{\DD} \right]_{2
\overline{2}} & \ldots & \left[ \mathbf{G}^{\DD} \right]_{2
\overline{L}} \\ \vdots & \vdots & \ddots & \vdots & \vdots &
\vdots & \ddots & \vdots \\ \left[ \mathbf{G}^{\DD} \right]_{L
1} & \left[ \mathbf{G}^{\DD} \right]_{L 2} & \ldots & \left[
\mathbf{G}^{\DD} \right]_{L L} & \left[ \mathbf{G}^{\DD}
\right]_{L \overline{1}} & \left[ \mathbf{G}^{\DD} \right]_{L
\overline{2}} & \ldots & \left[ \mathbf{G}^{\DD} \right]_{L
\overline{L}} \\ \hline \left[ \mathbf{G}^{\DD}
\right]_{\overline{1} 1} & \left[ \mathbf{G}^{\DD}
\right]_{\overline{1} 2} & \ldots & \left[ \mathbf{G}^{\DD}
\right]_{\overline{1} L} & \left[ \mathbf{G}^{\DD}
\right]_{\overline{1} \overline{1}} & \left[ \mathbf{G}^{\DD}
\right]_{\overline{1} \overline{2}} & \ldots & \left[
\mathbf{G}^{\DD} \right]_{\overline{1} \overline{L}} \\ \left[
\mathbf{G}^{\DD} \right]_{\overline{2} 1} & \left[
\mathbf{G}^{\DD} \right]_{\overline{2} 2} & \ldots & \left[
\mathbf{G}^{\DD} \right]_{\overline{2} L} & \left[
\mathbf{G}^{\DD} \right]_{\overline{2} \overline{1}} & \left[
\mathbf{G}^{\DD} \right]_{\overline{2} \overline{2}} & \ldots &
\left[ \mathbf{G}^{\DD} \right]_{\overline{2} \overline{L}} \\
\vdots & \vdots & \ddots & \vdots & \vdots & \vdots & \ddots &
\vdots \\ \left[ \mathbf{G}^{\DD} \right]_{\overline{L} 1} &
\left[ \mathbf{G}^{\DD} \right]_{\overline{L} 2} & \ldots &
\left[ \mathbf{G}^{\DD} \right]_{\overline{L} L} & \left[
\mathbf{G}^{\DD} \right]_{\overline{L} \overline{1}} & \left[
\mathbf{G}^{\DD} \right]_{\overline{L} \overline{2}} & \ldots &
\left[ \mathbf{G}^{\DD} \right]_{\overline{L} \overline{L}}
\end{array} \right) ,
\end{equation}
and similarly for \smash{$\mathbf{W}^{\DD}$} and
\smash{$\mathbf{\Sigma}^{\DD}$} (to be defined in a moment). For
the sake of simplicity, we shall disregard some subscripts and symbols
of dependence on $w$, $\overline{w}$.


We are interested in computing the Green's function of
\smash{$\widetilde{\mathbf{P}}$}, \ie
\begin{equation}\label{eq:EIG11}
G_{\widetilde{\mathbf{P}}} ( w , \overline{w} ) =
\frac{1}{N_{\tot}} \Tr \mathbf{G}^{w w} = \frac{1}{N_{\tot}}
\sum_{l = 1}^{L} \Tr \left[ \mathbf{G}^{\DD} \right]_{l l} =
\frac{1}{N_{\tot}} \sum_{l = 1}^{L} N_{l} \mathcal{G}_{l l} ,
\end{equation}
where it is useful to define the normalized traces
\begin{equation}\label{eq:EIG12}
\mathcal{G}_{l l} \equiv \frac{1}{N_{l}} \Tr \left[
\mathbf{G}^{\DD} \right]_{l l} , \qquad \mathcal{G}_{l
\overline{l}} \equiv \frac{1}{N_{l}} \Tr \left[ \mathbf{G}^{\DD}
\right]_{l \overline{l}} , \qquad\mathcal{G}_{\overline{l} l}
\equiv \frac{1}{N_{l}} \Tr \left[ \mathbf{G}^{\DD}
\right]_{\overline{l} l} , \qquad \mathcal{G}_{\overline{l}
\overline{l}} \equiv \frac{1}{N_{l}} \Tr \left[ \mathbf{G}^{\DD}
\right]_{\overline{l} \overline{l}} .
\end{equation}
Hence, we should evaluate the \smash{$\mathcal{G}_{l l}$}'s.


The only non--zero propagators of
\smash{$\widetilde{\mathbf{P}}^{\DD}$} are readily determined
from the probability measures in
(\ref{eq:RectangularGGMeasure}):
\begin{align}
\la \left[ \widetilde{\mathbf{P}}^{\DD} \right]_{1 2} \left[
\widetilde{\mathbf{P}}^{\DD} \right]_{\overline{2} \overline{1}}
\ra & = \frac{\sigma_{1}^{2}}{\sqrt{N_{1} N_{2}}} \Id_{N_{1}}
\otimes \Id_{N_{2}} , \nonumber\\
\la \left[ \widetilde{\mathbf{P}}^{\DD} \right]_{2 3} \left[
\widetilde{\mathbf{P}}^{\DD} \right]_{\overline{3} \overline{2}}
\ra & = \frac{\sigma_{2}^{2}}{\sqrt{N_{2} N_{3}}} \Id_{N_{2}}
\otimes \Id_{N_{3}} , \nonumber\\
& \vdots \nonumber\\
\la \left[ \widetilde{\mathbf{P}}^{\DD} \right]_{L 1} \left[
\widetilde{\mathbf{P}}^{\DD} \right]_{\overline{1} \overline{L}}
\ra & = \frac{\sigma_{L}^{2}}{\sqrt{N_{L} N_{1}}} \Id_{N_{L}}
\otimes \Id_{N_{1}}. \label{eq:EIG14}
\end{align}


Thus, we are now in position to write down the two
Dyson--Schwinger equations for
\smash{$\widetilde{\mathbf{P}}^{\DD}$}. The first one, being the
definition of the self--energy matrix $\mathbf{\Sigma}^{\DD}$, is independent 
of the propagators \cite{gjjn}:
\begin{equation}\label{eq:EIG16}
\mathbf{G}^{\DD} = \left( \mathbf{W}^{\DD} -
\mathbf{\Sigma}^{\DD} \right)^{- 1},
\end{equation}
where $\mathbf{W}^{\DD}$ is defined as $w
\Id_{N_{\mathrm{tot.}}}$ in its left upper block (where $w \in
\mathbb{C}$), $\overline{w} \Id_{N_{\mathrm{tot.}}}$ in the
right lower block, and zero elsewhere.
The second one is presented in \cite{gjjn,bjw}, and the structure of
the propagators (\ref{eq:EIG14}) implies that the only non--zero
blocks of the self--energy matrix read
\begin{equation}\label{eq:EIG17}
\left[ \mathbf{\Sigma}^{\DD} \right]_{l \overline{l}} =
\frac{\sigma_{l}^{2}}{\sqrt{N_{l} N_{l + 1}}} \Tr \left[
\mathbf{G}^{\DD} \right]_{l + 1 , \overline{l + 1}} \Id_{N_{l}}
= \underbrace{\sigma_{l}^{2} \sqrt{\frac{N_{l + 1}}{N_{l}}}
\mathcal{G}_{l + 1 , \overline{l + 1}}}_{\equiv \alpha_{l}}
\Id_{N_{l}} ,
\end{equation}
\begin{equation}\label{eq:EIG18}
\left[ \mathbf{\Sigma}^{\DD} \right]_{\overline{l} l} =
\frac{\sigma_{l - 1}^{2}}{\sqrt{N_{l - 1} N_{l}}} \Tr \left[
\mathbf{G}^{\DD} \right]_{\overline{l - 1} , l - 1} \Id_{N_{l}}
= \underbrace{\sigma_{l - 1}^{2} \sqrt{\frac{N_{l - 1}}{N_{l}}}
\mathcal{G}_{\overline{l - 1} , l - 1}}_{\equiv \beta_{l}}
\Id_{N_{l}} ,
\end{equation}
for all $l = 1 , 2 , \ldots , L$, with the cyclic convention $0
= L$, where the normalized traces (\ref{eq:EIG12}) have been
used.


Results (\ref{eq:EIG17}), (\ref{eq:EIG18}) mean that the four
blocks of the matrix \smash{$( \mathbf{W}^{\DD} -
\mathbf{\Sigma}^{\DD} )$} are diagonal. Such a matrix can be
straightforwardly inverted: its four blocks remain diagonal, and
read
\begin{equation}\label{eq:EIG20}
\left( \mathbf{W}^{\DD} - \mathbf{\Sigma}^{\DD} \right)^{- 1} =
\left( \begin{array}{cccc|cccc} \overline{w} \gamma_{1}
\Id_{N_{1}} & \Zero & \ldots & \Zero & \alpha_{1} \gamma_{1}
\Id_{N_{1}} & \Zero & \ldots & \Zero \\ \Zero & \overline{w}
\gamma_{2} \Id_{N_{2}} & \ldots & \Zero & \Zero & \alpha_{2}
\gamma_{2} \Id_{N_{2}} & \ldots & \Zero \\ \vdots & \vdots &
\ddots & \vdots & \vdots & \vdots & \ddots & \vdots \\ \Zero &
\Zero & \ldots & \overline{w} \gamma_{L} \Id_{N_{L}} & \Zero &
\Zero & \ldots & \alpha_{L} \gamma_{L} \Id_{N_{L}} \\ \hline
\beta_{1} \gamma_{1} \Id_{N_{1}} & \Zero & \ldots & \Zero & w
\gamma_{1} \Id_{N_{1}} & \Zero & \ldots & \Zero \\ \Zero &
\beta_{2} \gamma_{2} \Id_{N_{2}} & \ldots & \Zero & \Zero & w
\gamma_{2} \Id_{N_{2}} & \ldots & \Zero \\ \vdots & \vdots &
\ddots & \vdots & \vdots & \vdots & \ddots & \vdots \\ \Zero &
\Zero & \ldots & \beta_{L} \gamma_{L} \Id_{N_{L}} & \Zero &
\Zero & \ldots & w \gamma_{L} \Id_{N_{L}} \end{array} \right) ,
\end{equation}
where, for all $l = 1 , 2 , \ldots , L$,
\begin{equation}\label{eq:EIG21}
\frac{1}{\gamma_{l}} \equiv | w |^{2} - \alpha_{l} \beta_{l} = |
w |^{2} - \left( \sigma_{l - 1} \sigma_{l} \right)^{2}
\frac{\sqrt{N_{l - 1} N_{l + 1}}}{N_{l}} \mathcal{G}_{l + 1 ,
\overline{l + 1}} \mathcal{G}_{\overline{l - 1} , l - 1} .
\end{equation}
Substituting (\ref{eq:EIG20}) into (\ref{eq:EIG16}), we find
that the only non--zero blocks of the duplicated Green's
function (\ref{eq:EIG10}) are, for all $l = 1 , 2 , \ldots , L$,
\begin{equation}\label{eq:EIG22}
\left[ \mathbf{G}^{\DD} \right]_{l l} = \overline{w} \gamma_{l}
\Id_{N_{l}} , \qquad \left[ \mathbf{G}^{\DD} \right]_{l
\overline{l}} = \alpha_{l} \gamma_{l} \Id_{N_{l}} , \qquad
\left[ \mathbf{G}^{\DD} \right]_{\overline{l} l} = \beta_{l}
\gamma_{l} \Id_{N_{l}} , \qquad \left[ \mathbf{G}^{\DD}
\right]_{\overline{l} \overline{l}} = w \gamma_{l} \Id_{N_{l}} .
\end{equation}
Taking the normalized traces of both sides of every equality in
(\ref{eq:EIG22}) leads to the final set of equations,
\begin{equation}\label{eq:EIG23}
\mathcal{G}_{l l} = \overline{w} \gamma_{l} , \qquad
\mathcal{G}_{l \overline{l}} = \alpha_{l} \gamma_{l} , \qquad
\mathcal{G}_{\overline{l} l} = \beta_{l} \gamma_{l} , \qquad
\mathcal{G}_{\overline{l} \overline{l}} = w \gamma_{l} .
\end{equation}


The structure of equations (\ref{eq:EIG23}) is the following:
the fourth one is the conjugate of the first, and it is then
redundant. The second and third ones read
\begin{equation}\label{eq:EIG24}
\mathcal{G}_{l \overline{l}} = \sigma_{l}^{2} \sqrt{\frac{N_{l +
1}}{N_{l}}} \mathcal{G}_{l + 1 , \overline{l + 1}} \gamma_{l} ,
\end{equation}
\begin{equation}\label{eq:EIG25}
\mathcal{G}_{\overline{l} l} = \sigma_{l - 1}^{2}
\sqrt{\frac{N_{l - 1}}{N_{l}}} \mathcal{G}_{\overline{l - 1} , l
- 1} \gamma_{l} .
\end{equation}
We see that (\ref{eq:EIG21}), (\ref{eq:EIG24}) and
(\ref{eq:EIG25}) form a closed set of $3 L$ equations for $3 L$
unknowns, \smash{$\mathcal{G}_{l \overline{l}}$},
\smash{$\mathcal{G}_{\overline{l} l}$} and \smash{$\gamma_{l}$}.
Once solved, when the \smash{$\gamma_{l}$}'s have been found, we are
able to recover the Green's function of
\smash{$\widetilde{\mathbf{P}}$}, and subsequently the
$M$--transforms of \smash{$\widetilde{\mathbf{P}}$} and
$\mathbf{P}$ (in the argument \smash{$w^{L}$})
(\ref{eq:EIG07a}),
\begin{equation}\label{eq:EIG26}
G_{\widetilde{\mathbf{P}}} ( w , \overline{w} ) = \overline{w}
\frac{1}{N_{\tot}} \sum_{l = 1}^{L} N_{l} \gamma_{l} , \qquad
\textrm{\ie} \qquad M_{\widetilde{\mathbf{P}}} ( w ,
\overline{w} ) = \frac{1}{N_{\tot}} \sum_{l = 1}^{L} N_{l}
\mu_{l} , \qquad \textrm{\ie} \qquad M_{\mathbf{P}} \left( w^{L}
, \overline{w}^{L} \right) = \frac{1}{L} \sum_{l = 1}^{L} R_{l}
\mu_{l} ,
\end{equation}
where we have traded the \smash{$\gamma_{l}$}'s for a more
convenient set of variables,
\begin{equation}\label{eq:EIG27}
\mu_{l} \equiv | w |^{2} \gamma_{l} - 1.
\end{equation}


Equations (\ref{eq:EIG24}) and (\ref{eq:EIG25}) form a set of
decoupled recurrence relations for \smash{$\mathcal{G}_{l
\overline{l}}$} and \smash{$\mathcal{G}_{\overline{l} l}$},
respectively. Iterating these recurrences down to $l = 1$ gives
us:
\begin{equation}\label{eq:EIG28}
\mathcal{G}_{l \overline{l}} = \mathcal{G}_{1 \overline{1}}
\frac{1}{\left( \sigma_{1} \sigma_{2} \ldots \sigma_{l - 1}
\right)^{2}} \frac{1}{\sqrt{R_{l}}} \frac{1}{\gamma_{1}
\gamma_{2} \ldots \gamma_{l - 1}} ,
\end{equation}
\begin{equation}\label{eq:EIG29}
\mathcal{G}_{\overline{l} l} = \mathcal{G}_{\overline{1} 1}
\left( \sigma_{1} \sigma_{2} \ldots \sigma_{l - 1} \right)^{2}
\frac{1}{\sqrt{R_{l}}} \gamma_{2} \ldots \gamma_{l} .
\end{equation}
Applying the cyclic convention $0=L$, we get the following
equation:
\begin{equation}\label{eq:EIG30}
\left( \sigma_{1} \sigma_{2} \ldots \sigma_{L} \right)^{2}
\left( \gamma_{1} \gamma_{2} \ldots \gamma_{L} \right)
\mathcal{G}_{1 \overline{1}} = \mathcal{G}_{1 \overline{1}}.
\end{equation}
Straightforwardly, we get a trivial solution:
\smash{$\mathcal{G}_{l \overline{l}} = 0$} for all $l$, \ie
remembering (\ref{eq:EIG21}), \smash{$\gamma_{l} = 1 / | w
|^{2}$}, or equivalently \smash{$\mu_{l} = 0$} from (\ref{eq:EIG27}), and
therefore \smash{$M_{\mathbf{P}} ( z ,\overline{z} ) = 0$} 
from (\ref{eq:EIG26}). This is the holomorphic solution, holding
outside the eigenvalue density domain. In order to retrieve
information on the eigenvalue distribution, let us take
\smash{$\mathcal{G}_{1 \overline{1}} \neq 0$}.


After a change of variables to $\mu_{l}$ and some
simplifications, (\ref{eq:EIG21}) becomes
\begin{equation}\label{eq:EIG32}
\mu_{l} = \frac{| w |^{2} \mathcal{G}_{1 \overline{1}}
\mathcal{G}_{\overline{1} 1}}{\mu_{1} + 1} \frac{1}{R_{l}} ,
\end{equation}
from which we get the relation:
\begin{equation}\label{eq:EIG34}
\frac{\mu_{l}}{R_{1}}=\mu_{l} = \frac{\mu_{1}}{R_{l}} .
\end{equation}
After plugging (\ref{eq:EIG34}) into (\ref{eq:EIG30}), in terms of
the $\mu_{l}$ variables we obtain:
\begin{equation}\label{eq:EIG35}
\left( \mu_{1} + 1 \right) \left( \frac{\mu_{1}}{R_{2}} + 1
\right) \ldots \left( \frac{\mu_{1}}{R_{L}} + 1 \right) =
\frac{\left| w^{L} \right|^{2}}{\sigma^{2}} .
\end{equation}


On the other hand, substituting (\ref{eq:EIG34}) into
(\ref{eq:EIG26}), we get
\begin{equation}\label{eq:EIG36}
M_{\mathbf{P}} \left( w^{L} , \overline{w}^{L} \right) = \mu_{1}
.
\end{equation}
All in all, after changing the argument from $w$ to \smash{$z =
w^{L}$} we see that \smash{$M_{\mathbf{P}} ( z , \overline{z}
)$} obeys the $L$--th order polynomial equation
\begin{equation}\label{eq:EIG37}
\left( \frac{M_{\mathbf{P}} ( z , \overline{z} )}{R_{1}} + 1
\right) \left( \frac{M_{\mathbf{P}} ( z , \overline{z} )}{R_{2}}
+ 1 \right) \ldots \left( \frac{M_{\mathbf{P}} ( z ,
\overline{z} )}{R_{L}} + 1 \right) = \frac{| z
|^{2}}{\sigma^{2}} ,
\end{equation}
which is precisely the first main result of our article,
(\ref{eq:MPBasicEquation}).


The last point to be addressed is to determine the validity
domain of the non--holomorphic solution (\ref{eq:EIG37}),
knowing \cite{jnpz1} that on the boundary of such a domain, the
non--holomorphic and holomorphic solutions must be joined. Thus,
plugging the latter (\smash{$M_{\mathbf{P}} ( z , \overline{z} )
= 0$}) into (\ref{eq:EIG37}), we obtain an equation for the
borderline:
\begin{equation}\label{eq:EIG38}
| z | = \sigma .
\end{equation}
This means that the eigenvalues of the $\mathbf{P}$ matrix are
scattered on average, with the density stemming from
(\ref{eq:EIG37}), within a centered circle of radius $\sigma$.


When $L=2$, (\ref{eq:EIG37}) is just a second degree equation,
and it is easily solved. Indeed, in this case the
non--holomorphic $M$--transform reads
\begin{equation}\label{eq:EIG39}
M_{\mathbf{P}} ( z , \overline{z} ) = \frac{1}{2} \left( - 1 - R
+ \sqrt{( 1 - R )^{2} + 4 R \frac{| z |^{2}}{\sigma^{2}}}
\right) ,
\end{equation}
where we pose \smash{$R \equiv R_{2} = N_{2} / N_{1}$}, and
where the proper solution of (\ref{eq:EIG37}) has been picked up
in order to satisfy the matching condition (\ref{eq:EIG38}) with the 
holomorphic one on the borderline. As a result, we
immediately obtain the Green's function:
\begin{equation}\label{eq:EIG40}
G_{\mathbf{P}} ( z , \overline{z} ) = \frac{1}{2 z} \left( 1 - R
+ \sqrt{( 1 - R )^{2} + 4 R \frac{| z |^{2}}{\sigma^{2}}}
\right).
\end{equation}

When deriving the average spectral density, one has to be
cautious in the vicinity of the origin of the complex plane in
order to properly take possible zero modes into account. Let us
first expand (\ref{eq:EIG40}) near $z = 0$ in order to clarify
its behavior:
\begin{equation}\label{eq:EIG40a}
G_{\mathbf{P}} ( z , \overline{z} ) \sim \frac{f}{z} +
\textrm{regular terms} , \qquad \textrm{as} \qquad z \to 0 ,
\qquad \textrm{where} \qquad f \equiv \left\{ \begin{array}{ll}
1 - R , & \quad \textrm{for} \quad R < 1 , \\ 0 , & \quad
\textrm{for} \quad R \geq 1 . \end{array} \right.
\end{equation}

Taking the derivative \smash{$( 1 / \pi )
\partial_{\overline{z}}$} of this singular term yields a Dirac
delta function at the origin, \smash{$f \delta^{( 2 )} ( z ,
\overline{z} )$}. Altogether,
\begin{equation}\label{eq:EIG41}
\rho_{\mathbf{P}} ( z , \overline{z} ) = \left\{
\begin{array}{ll} \frac{1}{\pi \sigma^{2}} \frac{R}{\sqrt{( 1 -
R )^{2} + 4 R \frac{| z |^{2}}{\sigma^{2}}}} + f \delta^{( 2 )}
( z , \overline{z} ) , & \quad \textrm{for} \quad | z | \leq
\sigma , \\ 0 , & \quad \textrm{for} \quad | z | > \sigma .
\end{array} \right.
\end{equation}

\begin{figure}[h]
\includegraphics[width=8.5cm]{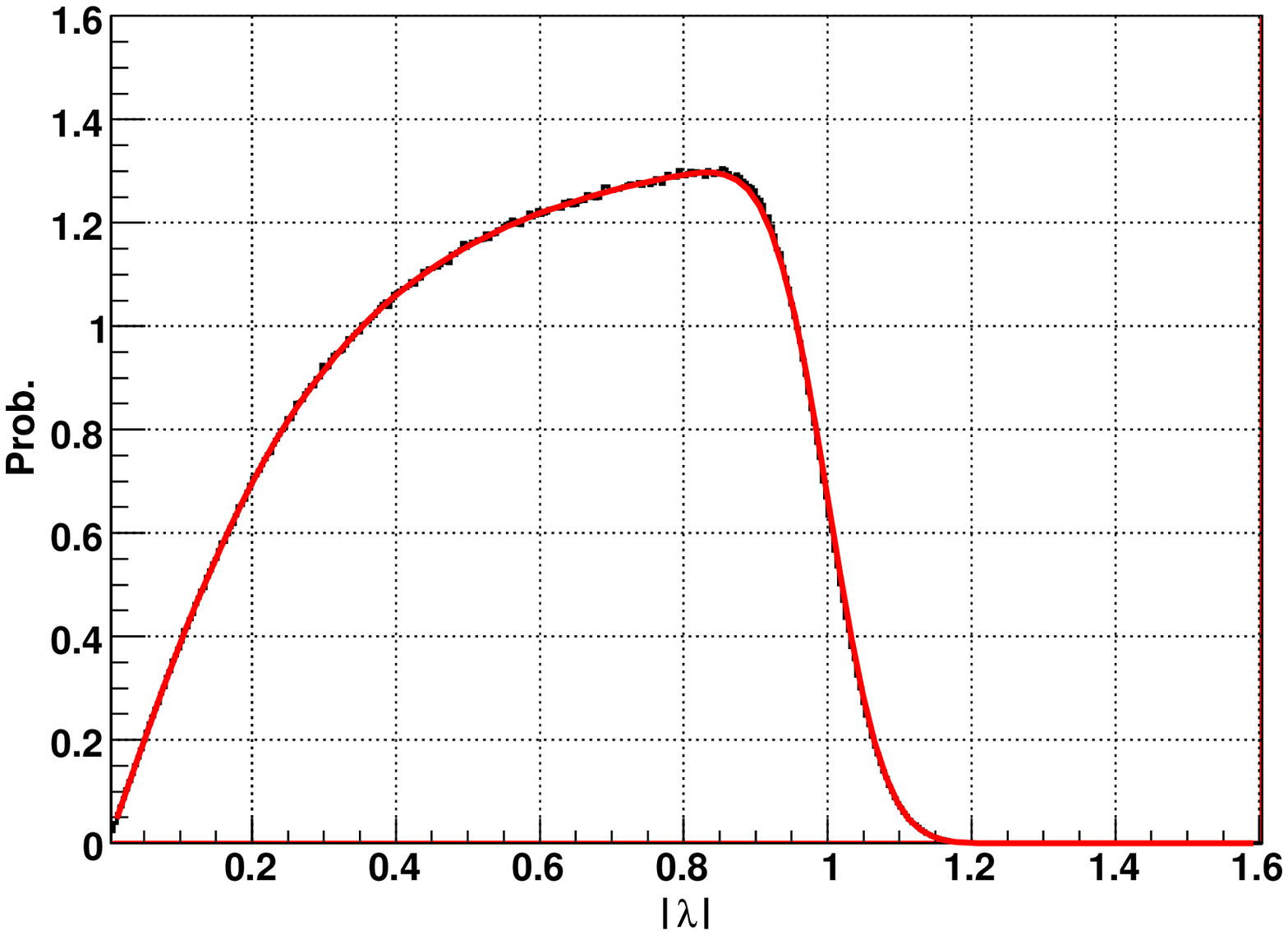}
\includegraphics[width=8.5cm]{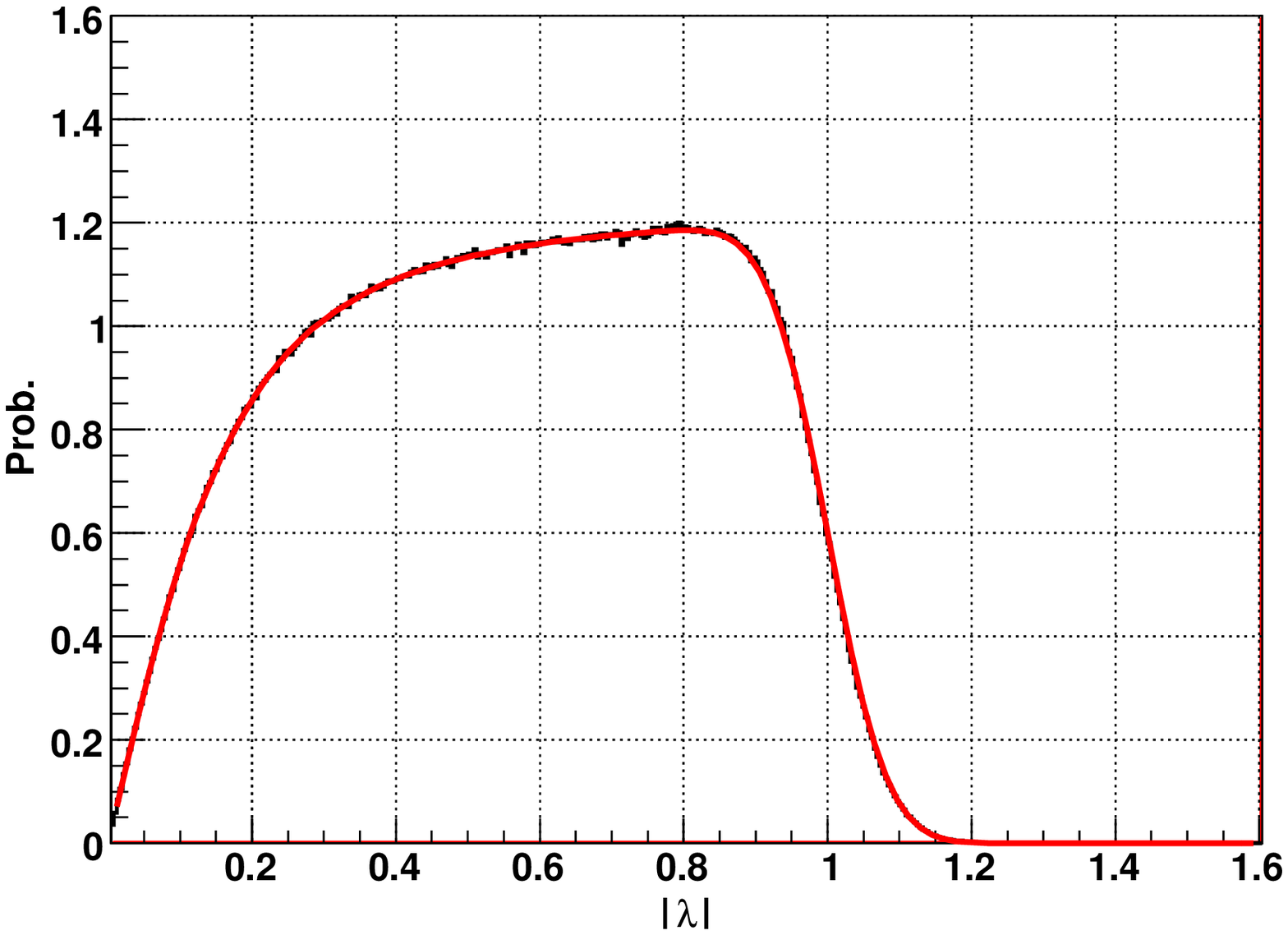}
\includegraphics[width=8.5cm]{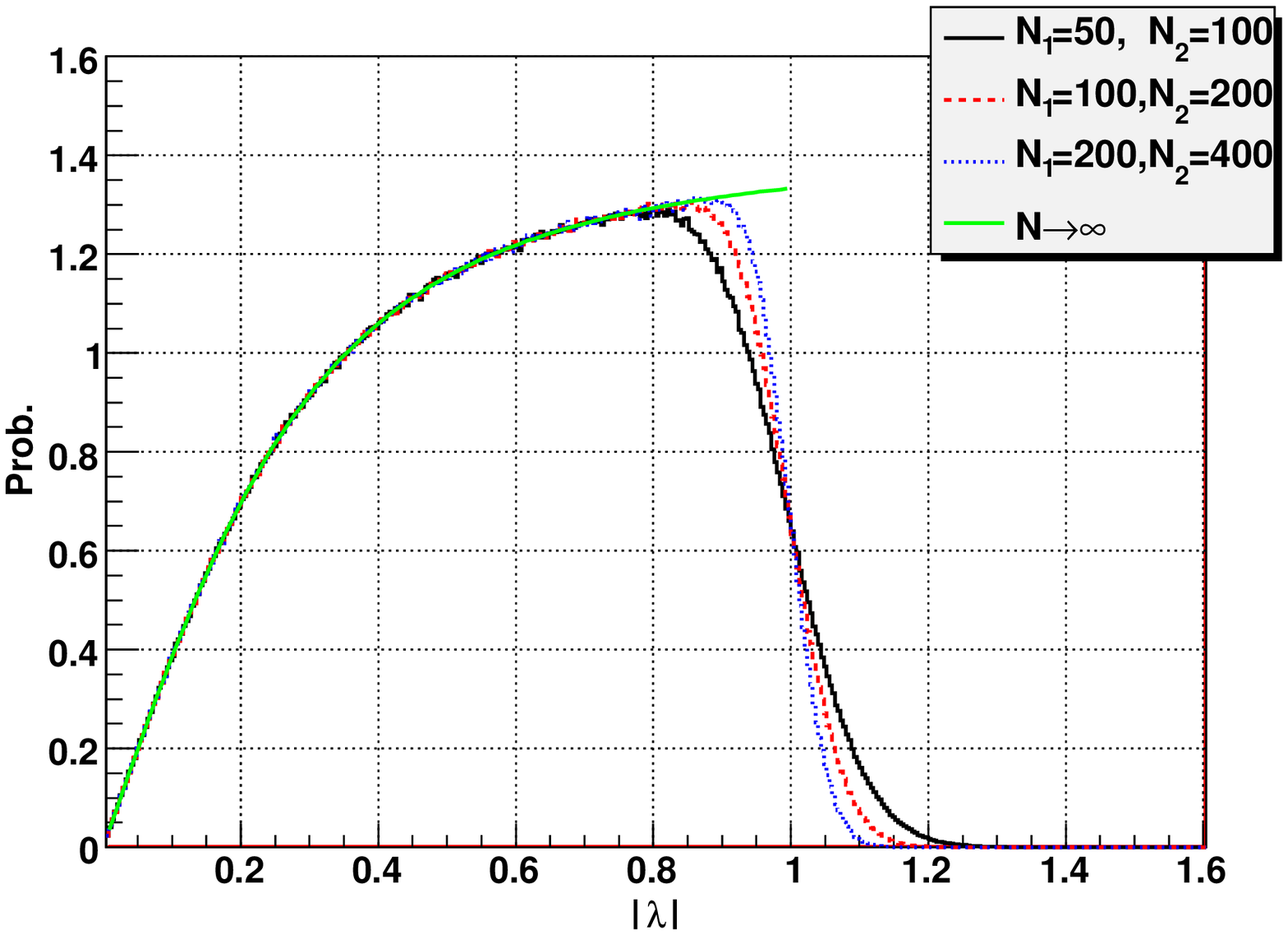}
\includegraphics[width=8.5cm]{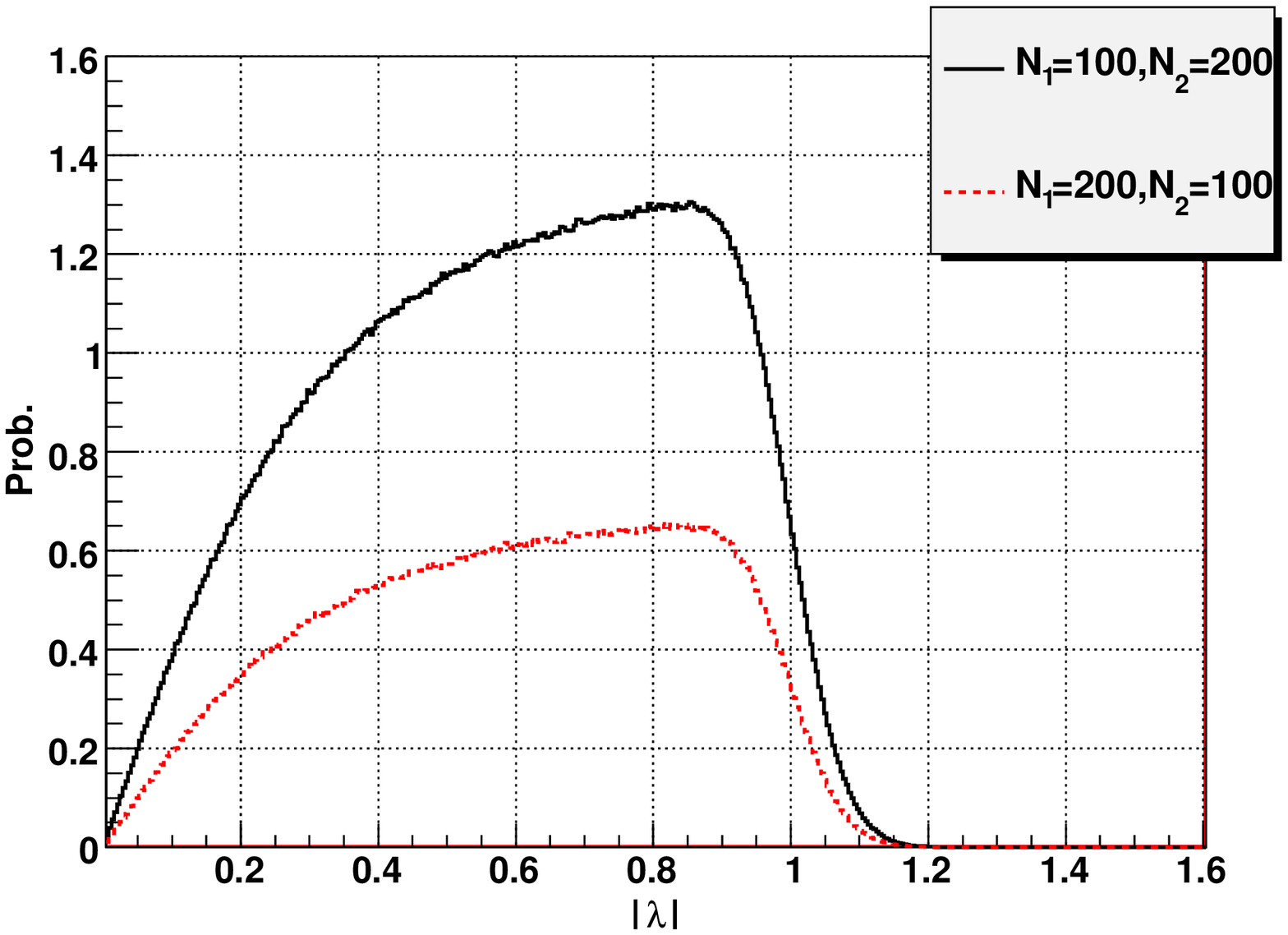}
\caption{Numerical verification of the theoretical formula
(\ref{eq:EIG41}) for (the radial part (\ref{eq:EIG46}) of) the
mean spectral density \smash{$\rho_{\mathbf{P}} ( z ,
\overline{z} )$} of the product $\mathbf{P}$ of $L = 2$
rectangular Gaussian random matrices, as well as the
finite--size correction (\ref{eq:EIG47}).\\
UP LEFT: A numerical histogram (the black line) versus the
theoretical prediction (\ref{eq:EIG41}), supplemented with the
finite--size smoothing (\ref{eq:EIG47}) (the red plot), for
\smash{$N_{1} = 100$} and \smash{$N_{2} = 200$} (\ie \smash{$R =
R_{2} = 2$}), and for \smash{$10^{5}$} Monte--Carlo iterations
(\ie the histogram is made of \smash{$10^{7}$} eigenvalues). The
adjustable parameter $q$ (\ref{eq:EIG47}) is fitted to be $q
\approx 1.14$.\\
UP RIGHT: An analogous graph to UP LEFT, this time with
\smash{$N_{1} = 100$} and \smash{$N_{2} = 150$} (\ie $R = 1.5$).
We find $q \approx 1.08$ here.\\
DOWN LEFT: An analysis of the finite--size effects: numerical
histograms for \smash{$N_{1} = 50$}, \smash{$N_{2} = 100$}
(black), \smash{$N_{1} = 100$}, \smash{$N_{2} = 200$} (dashed
red), \smash{$N_{1} = 200$}, \smash{$N_{2} = 400$} (dotted
blue), \ie with the same rectangularity ratio $R = 2$, but
increasing matrix dimensions. We observe how these plots
approach the green line of the theoretical formula
(\ref{eq:EIG41}) for the density in the thermodynamic limit.\\
DOWN RIGHT: Numerical histograms for the matrix sizes of
\smash{$N_{1} = 100$}, \smash{$N_{2} = 200$} (\ie $R = 2$;
black) and \smash{$N_{1} = 200$}, \smash{$N_{2} = 100$} (\ie $R
= 1 / 2$; red). Due to the presence of the zero modes (not
displayed in the picture), the latter is half of the former.}
\label{fig:EIGL2}
\end{figure}

\begin{figure}[h]
\includegraphics[width=8.5cm]{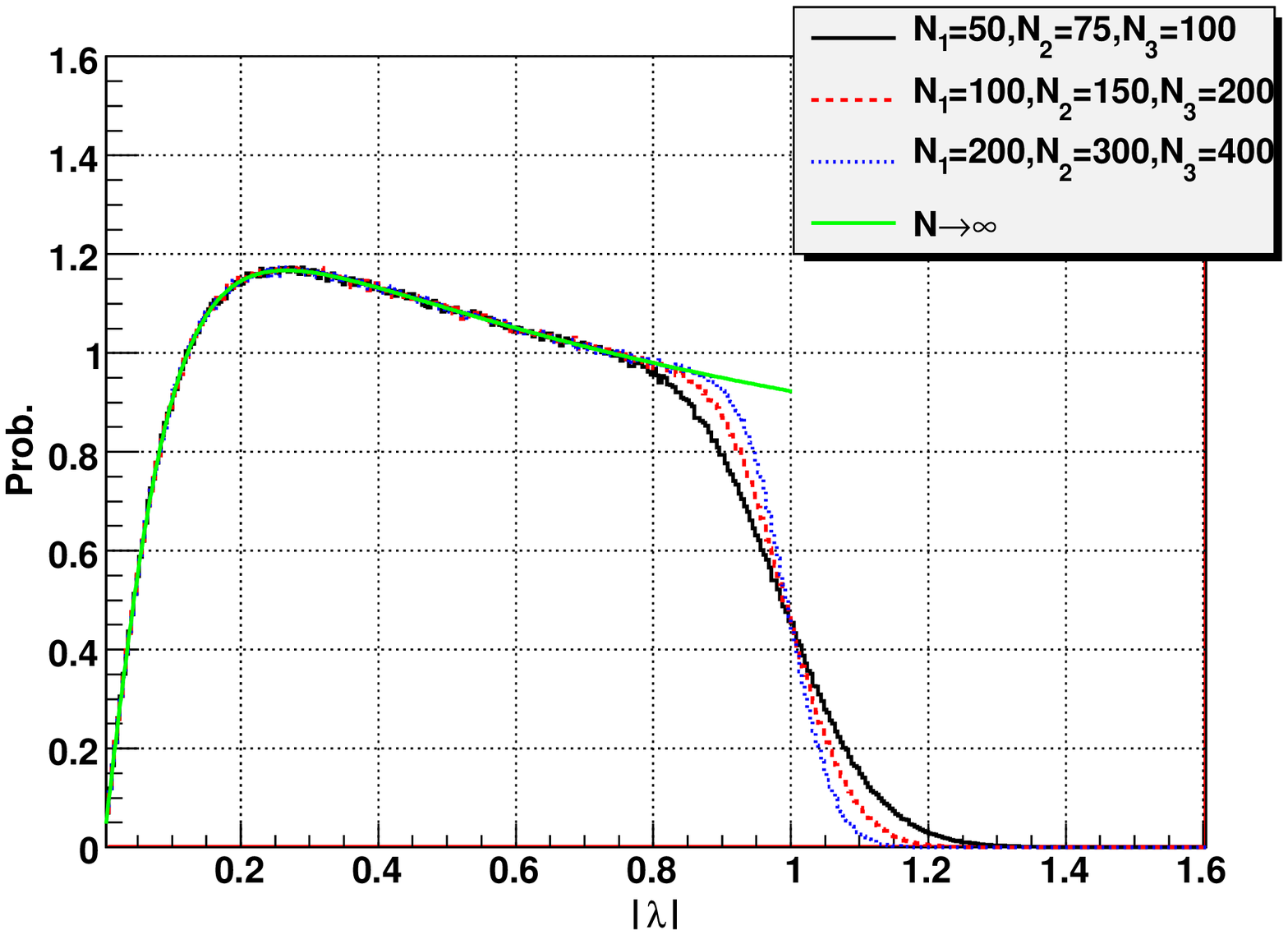}
\includegraphics[width=8.5cm]{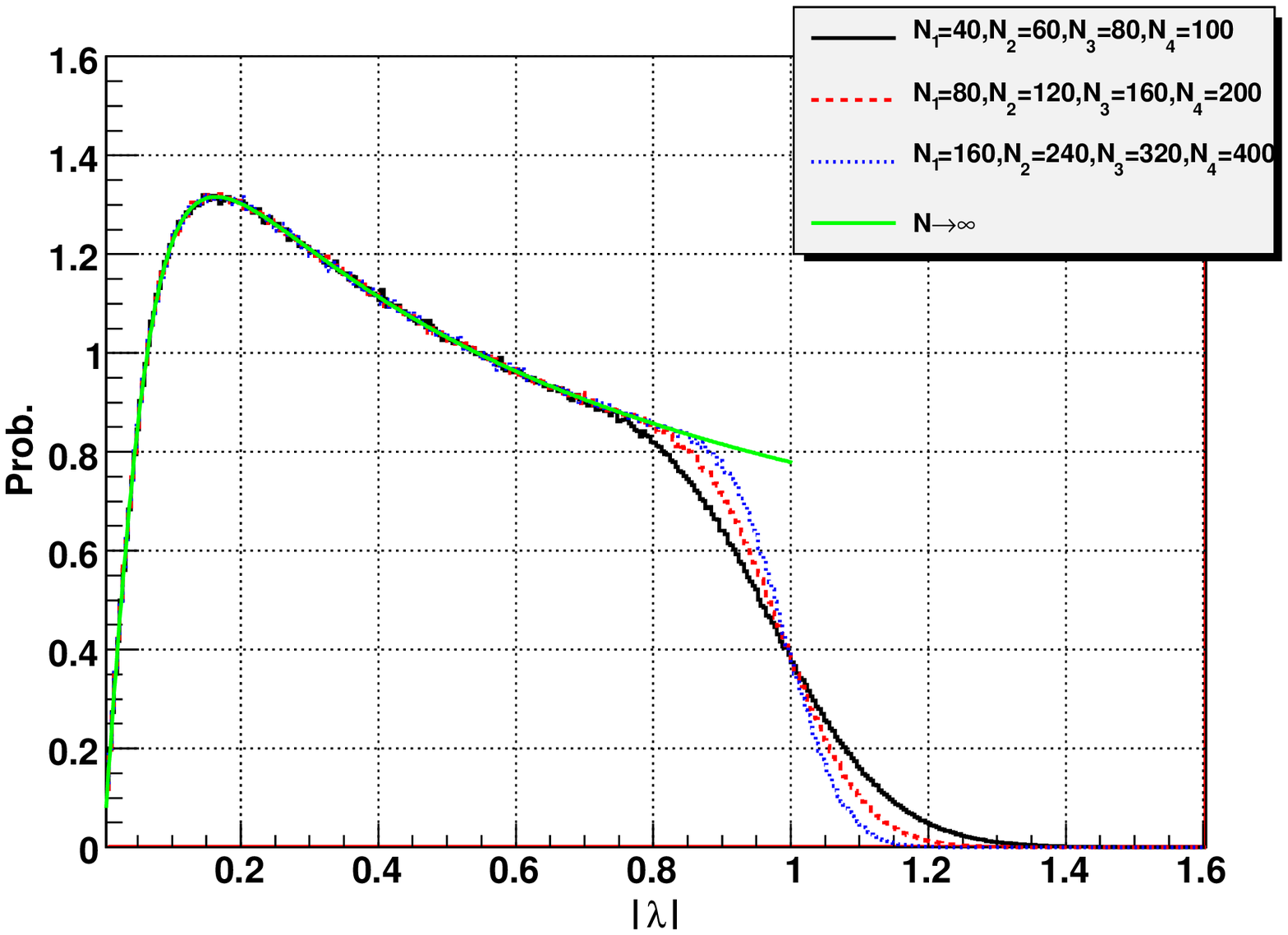}
\caption{Analogous graphs to figure~\ref{fig:EIGL2}, DOWN LEFT,
but for $L = 3$ (LEFT) and $L = 4$ (RIGHT).}
\label{fig:EIGL34}
\end{figure}

Moreover, one can also verify that the density, in the
thermodynamic limit, changes on the borderline from being
non--holomorphic with value given by
\begin{equation}\label{eq:EIG45}
\rho_{\mathbf{P}} ( z , \overline{z} ) \left|_{| z | = \sigma}
\right. = \frac{1}{\pi \sigma^{2}} R_{\mathrm{h}} , \qquad
\textrm{where} \qquad \frac{1}{R_{\mathrm{h}}} \equiv \sum_{l =
1}^{L} \frac{1}{R_{l}},
\end{equation}
to being holomorphic with value $0$. However, for finite sizes of
the random matrices, this step gets smoothed out. Let us then
consider the radial density,
\begin{equation}\label{eq:EIG46}
\rho_{\mathbf{P}}^{\textrm{rad.}} ( r ) \equiv 2 \pi r
\rho_{\mathbf{P}} ( z , \overline{z} ) \left|_{| z | = r}
\right.
\end{equation}
and, following \cite{bjw}, let us propose the following model for
this finite--$N$ effect (where by $N$ we denote the order of
magnitude of the dimensions of the matrices, say \smash{$N
\equiv N_{1}$}). We introduce an ``effective''
radial density in order to properly incorporate such finite--$N$
behavior at the borderline,
\begin{equation}\label{eq:EIG47}
\rho_{\mathbf{P}}^{\textrm{eff.}} ( r ) \equiv
\rho_{\mathbf{P}}^{\textrm{rad.}} ( r ) \frac{1}{2}
\mathrm{erfc} \left( q ( r - \sigma ) \sqrt{N} \right) ,
\end{equation}
where $q$ is a free parameter whose value is to be adjusted by
fitting. We numerically verify this hypothesis (see figures
\ref{fig:EIGL2} and \ref{fig:EIGL34}).

\section{The Singular Values of a Product of Rectangular
Gaussian Random Matrices}
\label{s:TheSingularValuesOfAProductOfRectangularGaussianRandomMatrices}


In the following we show how to derive formula
 (\ref{eq:MQBasicEquation}), \ie an $( L + 1
)$--th order polynomial equation obeyed by the $M$--transform
(which, as already discussed, encodes the same information
contained in the spectral density) of the Hermitian matrix
\smash{$\mathbf{Q} \equiv \mathbf{P}^{\dagger} \mathbf{P}$}
(\ref{eq:SingularValuesDefinition}), $\mathbf{P}$ being the
product (\ref{eq:ProductDefinition}) of rectangular
(\ref{eq:ThermodynamicLimit}) Gaussian random matrices
(\ref{eq:RectangularGGMeasure}). Our result agrees with that in~\cite{MULLER}, obtained in the context of wireless telecommunication theory, provided we
synchronize the conventions. In particular, our resolvent $G(z)$ relates to Stieltjes transform as $G(z)=-G(-s)$. The underlying idea will be to
rewrite $\mathbf{Q}$ as a product of some Hermitian matrices in
order to apply the techniques provided by Free Random Variables
(FRV) calculus. Loosely speaking, FRV calculus (initiated by the
pioneering works of Speicher and Voiculescu \emph{et al.}) can
be thought as the extension of standard probability theory to
non--commutative objects, such as matrices. Given the broadness
of the topic, we shall not attempt any introductory discussion
here, and we refer the non--expert reader to \cite{vdn, s}.

Let us commence by defining, for any $l = 1 , 2 , \ldots , L$,
a square \smash{$N_{l + 1} \times N_{l + 1}$} matrix
\begin{equation}\label{eq:SV01}
\mathbf{Q}_{l} \equiv \left( \mathbf{A}_{1} \mathbf{A}_{2}
\ldots \mathbf{A}_{l - 1} \mathbf{A}_{l} \right)^{\dagger}
\left( \mathbf{A}_{1} \mathbf{A}_{2} \ldots \mathbf{A}_{l - 1}
\mathbf{A}_{l} \right) = \mathbf{A}_{l}^{\dagger} \mathbf{A}_{l
- 1}^{\dagger} \ldots \mathbf{A}_{2}^{\dagger}
\mathbf{A}_{1}^{\dagger} \mathbf{A}_{1} \mathbf{A}_{2} \ldots
\mathbf{A}_{l - 1} \mathbf{A}_{l} ,
\end{equation}
being a generalization of $\mathbf{Q}$ which includes only the
first $l$ random matrices, as well as a square \smash{$N_{l}
\times N_{l}$} matrix, which differs from
\smash{$\mathbf{Q}_{l}$} only in the position of the last matrix
in the string, \ie \smash{$\mathbf{A}_{l}$}, which is now placed
as the first matrix in the string,
\begin{equation}\label{eq:SV02}
\widetilde{\mathbf{Q}}_{l} \equiv \mathbf{A}_{l}
\mathbf{A}_{l}^{\dagger} \mathbf{A}_{l - 1}^{\dagger} \ldots
\mathbf{A}_{2}^{\dagger} \mathbf{A}_{1}^{\dagger} \mathbf{A}_{1}
\mathbf{A}_{2} \ldots \mathbf{A}_{l - 1} = \left( \mathbf{A}_{l}
\mathbf{A}_{l}^{\dagger} \right) \mathbf{Q}_{l - 1} .
\end{equation}
We are interested in the eigenvalues of the Hermitian matrix
\smash{$\mathbf{Q} = \mathbf{Q}_{L}$}.

The orders of the terms in the two above products
(\ref{eq:SV01}), (\ref{eq:SV02}) are related to each other by a
cyclic shift, therefore, for any integer $n \geq 1$, there will
be \smash{$\Tr \mathbf{Q}_{l}^{n} = \Tr
\widetilde{\mathbf{Q}}_{l}^{n}$}. Hence, the $M$--transforms
(see equation \eqref{Mtrans}) of the two above random matrices
are related by the following relation
\begin{equation}\label{eq:SV03}
M_{\mathbf{Q}_{l}} ( z ) = \sum_{n \geq 1} \frac{1}{z^{n}}
\frac{1}{N_{l + 1}} \la \Tr \mathbf{Q}_{l}^{n} \ra =
\frac{N_{l}}{N_{l + 1}} \sum_{n \geq 1} \frac{1}{z^{n}}
\frac{1}{N_{l}} \la \Tr \widetilde{\mathbf{Q}}_{l}^{n} \ra =
\frac{R_{l}}{R_{l + 1}} M_{\widetilde{\mathbf{Q}}_{l}} ( z ).
\end{equation}
Now, let us consider the functional inverse of the
$M$--transform, called the $N$--transform, defined as:
$M_{\mathbf{Q}_l} (N_{\mathbf{Q}_l}(z)) = N_{\mathbf{Q}_l}
(M_{\mathbf{Q}_l} (z)) = z$. Employing this definition within
equation \eqref{eq:SV03} one easily obtains
\begin{equation}\label{eq:SV04}
N_{\mathbf{Q}_{l}} ( z ) = N_{\widetilde{\mathbf{Q}}_{l}} \left(
\frac{R_{l + 1}}{R_{l}} z \right) .
\end{equation}

Now, since it can be safely stated that independent random
matrices become free with respect to each other in the
thermodynamical limit, it becomes clear that the reason for
introducing the auxiliary matrix $\widetilde{\mathbf{Q}}_{l}$ is
that it is a product of two free matrices,
\smash{$\mathbf{A}_{l} \mathbf{A}_{l}^{\dagger}$} and
\smash{$\mathbf{Q}_{l - 1}$}. Then, the FRV multiplication
\cite{vdn} law for free matrices can be applied. Such law states
that the $N$--transform of the product of two free matrices,
$\mathbf{A}$ and $\mathbf{B}$, is simply given by $N_{\mathbf{A}
\mathbf{B}} (z) = z / (1+z) \ N_{\mathbf{A}} (z) N_{\mathbf{B}}
(z)$. (In the language more often found in the literature on the subject, the $N$--transform is replaced by the so--called $S$--transform, \smash{$S_{\mathbf{X}} ( z ) \equiv ( z + 1 ) / ( z N_{\mathbf{X}} ( z ) )$}, which then obeys a simpler multiplication law, \smash{$S_{\mathbf{A} \mathbf{B}} ( z ) = S_{\mathbf{A}} ( z ) S_{\mathbf{B}} ( z )$}). So, when applying this relation to the $\widetilde{\mathbf{Q}}_{l}$ matrix \eqref{eq:SV02} one can
write, for $l = 2 , 3 , \ldots , L$
\begin{equation}\label{eq:SV05}
N_{\widetilde{\mathbf{Q}}_{l}} ( z ) = \frac{z}{z + 1}
N_{\mathbf{A}_{l} \mathbf{A}_{l}^{\dagger}} ( z )
N_{\mathbf{Q}_{l - 1}} ( z ) .
\end{equation}

{}From equations (\ref{eq:SV04}) and (\ref{eq:SV05}), we now
eliminate the $N$--transform of the auxiliary
\smash{$\widetilde{\mathbf{Q}}_{l}$}, which leaves us with the
following recurrence relation for the $N$--transform of
\smash{$\mathbf{Q}_{l}$},
\begin{equation}\label{eq:SV06}
N_{\mathbf{Q}_{l}} ( z ) = \frac{z}{z + \frac{R_{l}}{R_{l + 1}}}
N_{\mathbf{A}_{l} \mathbf{A}_{l}^{\dagger}} \left( \frac{R_{l +
1}}{R_{l}} z \right) N_{\mathbf{Q}_{l - 1}} \left( \frac{R_{l +
1}}{R_{l}} z \right) , \qquad \textrm{for} \qquad l = 2 , 3 ,
\ldots , L ,
\end{equation}
with the initial condition,
\begin{equation}\label{eq:SV07}
N_{\mathbf{Q}_{1}} ( z ) = N_{\widetilde{\mathbf{Q}}_{1}} \left(
\frac{R_{2}}{R_{1}} z \right) = N_{\mathbf{A}_{1}
\mathbf{A}_{1}^{\dagger}} \left( \frac{R_{2}}{R_{1}} z \right) ,
\end{equation}
which stems from (\ref{eq:SV04}) and from the fact that
\smash{$\widetilde{\mathbf{Q}}_{1} = \mathbf{A}_{1}
\mathbf{A}_{1}^{\dagger}$}. The solution of this recurrence
(\ref{eq:SV06}), (\ref{eq:SV07}) is then readily found to be
\begin{equation}\label{eq:SV08}
N_{\mathbf{Q}_{L}} ( z ) = \frac{z^{L - 1}}{\left( z + R_{2}
\right) \left( z + R_{3} \right) \ldots \left( z + R_{L}
\right)} N_{\mathbf{A}_{1} \mathbf{A}_{1}^{\dagger}} \left(
\frac{z}{R_{1}} \right) N_{\mathbf{A}_{2}
\mathbf{A}_{2}^{\dagger}} \left( \frac{z}{R_{2}} \right) \ldots
N_{\mathbf{A}_{L} \mathbf{A}_{L}^{\dagger}} \left(
\frac{z}{R_{L}} \right) .
\end{equation}

It remains now to find the $N$--transforms of the random
matrices \smash{$\mathbf{A}_{l} \mathbf{A}_{l}^{\dagger}$}. They
are examples of the so--called ``Wishart ensembles'', and the
problem of computing their $N$--transforms, with the same
normalization of the probability measures
(\ref{eq:RectangularGGMeasure}) of the
\smash{$\mathbf{A}_{l}$}'s which we are employing, 
has first been solved in
\cite{fz}: expressions (1.8), (2.8), (2.13), (2.14) of this
article yield the Green's function of \smash{$\mathbf{A}_{l}
\mathbf{A}_{l}^{\dagger}$}, which immediately leads to the
pertinent $N$--transform,
\begin{equation}\label{eq:SV09}
N_{\mathbf{A}_{l} \mathbf{A}_{l}^{\dagger}} ( z ) =
\sigma_{l}^{2} \frac{( z + 1 ) \left( \sqrt{\frac{N_{l}}{N_{l +
1}}} z + \sqrt{\frac{N_{l + 1}}{N_{l}}} \right)}{z} .
\end{equation}
Substituting (\ref{eq:SV09}) into (\ref{eq:SV08}), one finally
arrives at the desired formula for the $N$--transform of
\smash{$\mathbf{Q} = \mathbf{Q}_{L}$},
\begin{equation}\label{eq:SV10}
N_{\mathbf{Q}} ( z ) = \sigma^{2} \sqrt{R_{1}} \frac{1}{z} ( z +
1 ) \left( \frac{z}{R_{1}} + 1 \right) \left( \frac{z}{R_{2}} +
1 \right) \ldots \left( \frac{z}{R_{L}} + 1 \right) ,
\end{equation}
with $\sigma$ defined as in the previous sections. In other
words, the corresponding $M$--transform \smash{$M_{\mathbf{Q}} (
z )$} satisfies the following polynomial equation of order $( L
+ 1 )$,
\begin{equation}\label{eq:SV11}
\sqrt{R_{1}} \frac{1}{M_{\mathbf{Q}} ( z )} \left(
M_{\mathbf{Q}} ( z ) + 1 \right) \left( \frac{M_{\mathbf{Q}} ( z
)}{R_{1}} + 1 \right) \left( \frac{M_{\mathbf{Q}} ( z )}{R_{2}}
+ 1 \right) \ldots \left( \frac{M_{\mathbf{Q}} ( z )}{R_{L}} + 1
\right) = \frac{z}{\sigma^{2}} ,
\end{equation}
or in the case of \smash{$N_{L + 1} = N_{1}$} (\ie \smash{$R_{1}
= 1$}, required when one wishes for $\mathbf{P}$ to have
eigenvalues too),
\begin{equation}\label{eq:SV12}
\frac{1}{M_{\mathbf{Q}} ( z )} \left( M_{\mathbf{Q}} ( z ) + 1
\right)^{2} \left( \frac{M_{\mathbf{Q}} ( z )}{R_{2}} + 1
\right) \ldots \left( \frac{M_{\mathbf{Q}} ( z )}{R_{L}} + 1
\right) = \frac{z}{\sigma^{2}} .
\end{equation}
This completes our derivation of (\ref{eq:MQBasicEquation}).

\begin{figure}[h]
\includegraphics[width=5.7cm]{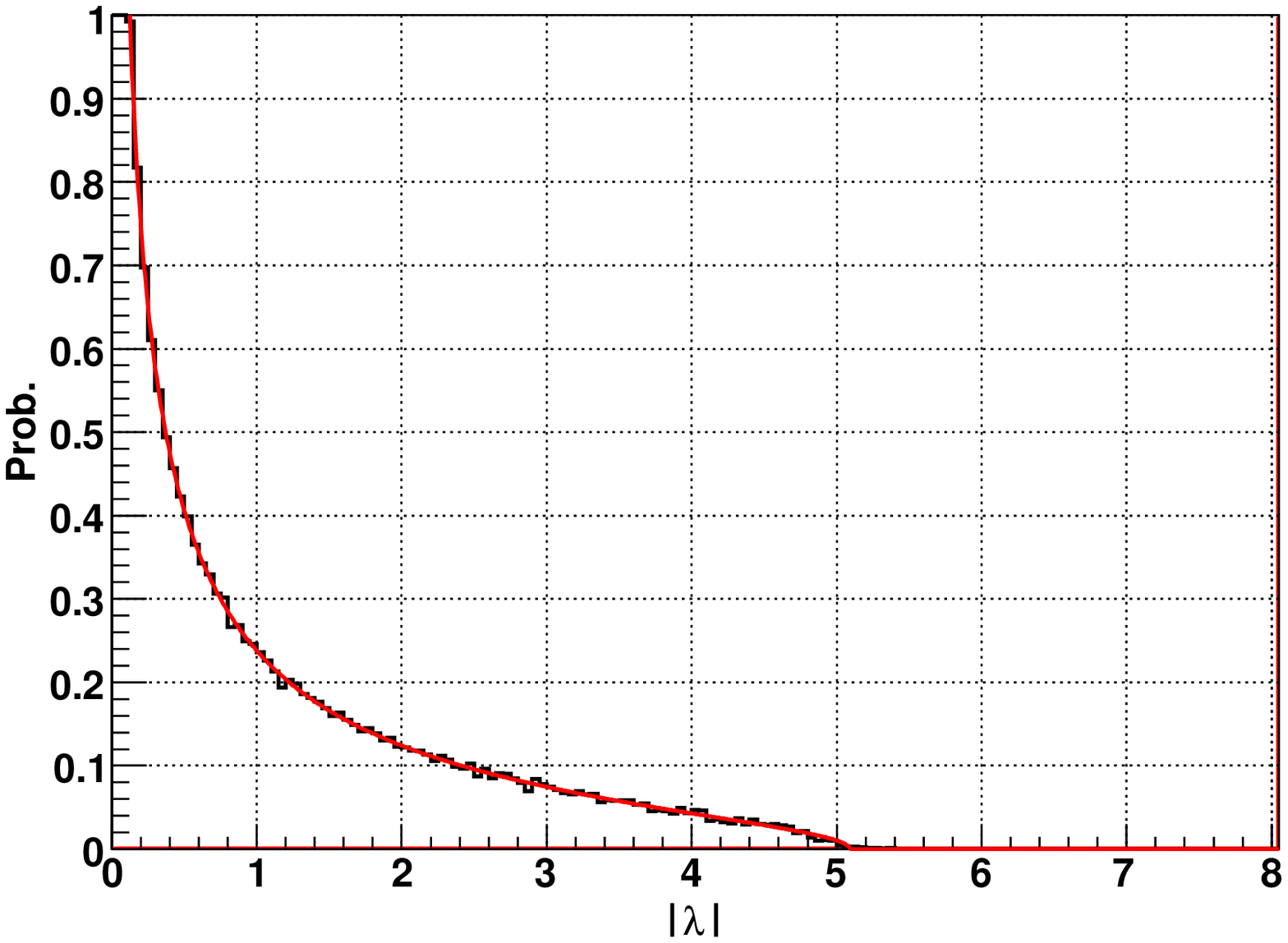}
\includegraphics[width=5.7cm]{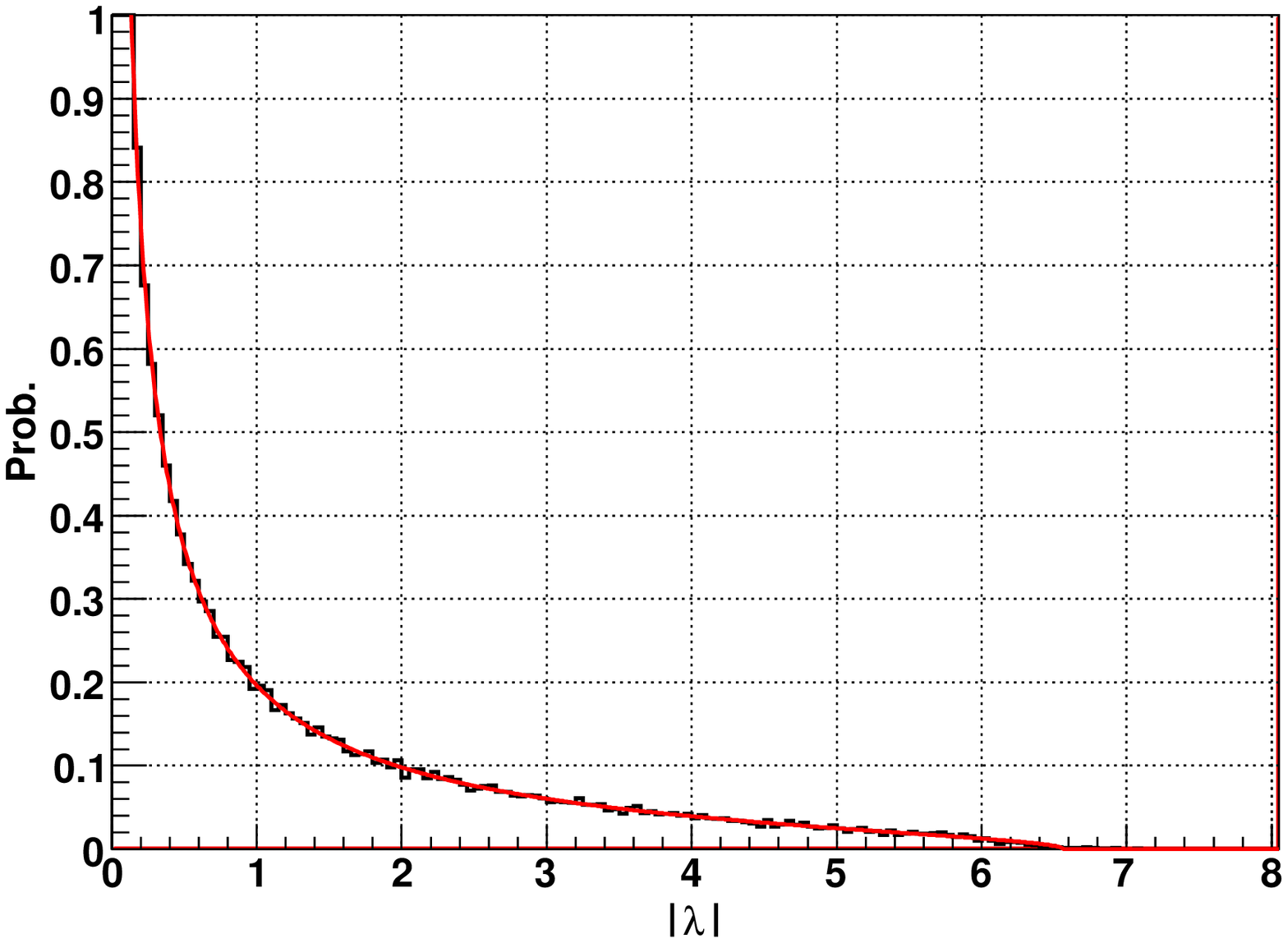}
\includegraphics[width=5.7cm]{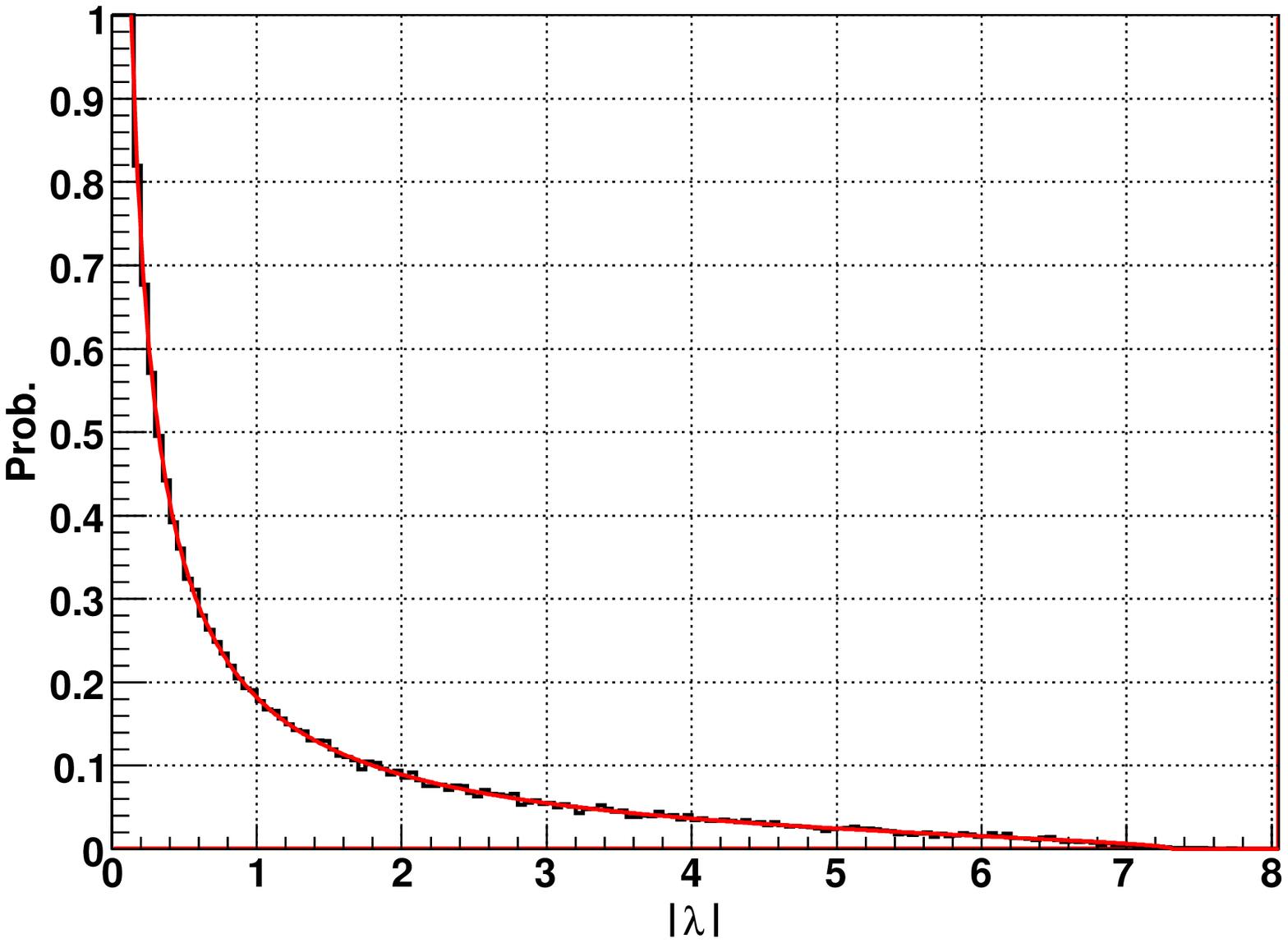}
\caption{Numerical verification of the theoretical formula
(\ref{eq:SV11}) for the mean spectral density
\smash{$\rho_{\mathbf{Q}} ( \lambda )$} of the random matrix
\smash{$\mathbf{Q} = \mathbf{P}^{\dagger} \mathbf{P}$}
(\ref{eq:SingularValuesDefinition}). Everywhere  we have \smash{$N_{L +
1} = N_{1} = 50$}. The number of Monte--Carlo iterations is
$20,\!000$, \ie all the histograms are generated from
\smash{$10^{6}$} eigenvalues.\\
LEFT: $L = 2$, and the matrix sizes are chosen to be
\smash{$N_{1} = 50$}, \smash{$N_{2} = 150$}.\\
MIDDLE: $L = 3$, and the matrix sizes are \smash{$N_{1} = 50$},
\smash{$N_{2} = 100$}, \smash{$N_{3} = 150$}.\\
RIGHT: $L = 4$, and the matrix sizes are \smash{$N_{1} = 50$},
\smash{$N_{2} = 100$}, \smash{$N_{3} = 150$}, \smash{$N_{4} =
200$}.}
\label{fig:SVL234}
\end{figure}

We have performed extended numerical tests of the formula
(\ref{eq:SV11}), in all cases obtaining perfect agreement,
see figure~\ref{fig:SVL234}.


\section{Conclusions}

The main contribution of this article is equation (\ref{eq:MPBasicEquation}) for the $M$--transforms of the product \smash{$\mathbf{P} = \mathbf{A}_{1} \mathbf{A}_{2} \ldots \mathbf{A}_{L}$} (\ref{eq:ProductDefinition}) of an
arbitrary number $L$ of independent rectangular
(\ref{eq:ThermodynamicLimit}) Gaussian random matrices
(\ref{eq:RectangularGGMeasure}). Knowing the $M$--transform one can easily
calculate the eigenvalue density of the product (\ref{rM}), which turns out
to be spherically symmetric in the complex plane. We also discussed
a striking resemblance of equation (\ref{eq:MPBasicEquation}) to
the corresponding equation (\ref{eq:MQBasicEquation})
of the Hermitian matrix \smash{$\mathbf{Q} = \mathbf{P}^{\dagger} \mathbf{P}$}
(\ref{eq:SingularValuesDefinition}), whose eigenvalues are equal to
the squared singular values of $\mathbf{P}$. Both
these equations are polynomial (of orders $L$ and $( L + 1 )$
respectively), so in general they may only be solved
numerically; however, some properties of the mean spectral
densities can still be retrieved analytically, such as their
singular behavior at zero (\ref{eq:RhoPSingularityAtZero}),
(\ref{eq:RhoQSingularityAtZero}).

We are tempted to conjecture
that this similarity of the $M$--transforms for $\mathbf{P}$ and
$\mathbf{Q}$ is generic for random matrices possessing rotationally
symmetric average distribution of the eigenvalues, and that the corresponding equations differ only by the prefactor which we have discussed while comparing
(\ref{eq:MPBasicEquation}) and (\ref{eq:MQBasicEquation}).
For such models, the non--holomorphic $M$--transform
\smash{$M_{\mathbf{X}} ( z , \overline{z} )$} is a function of
the real argument \smash{$| z |^{2}$}, thereby allowing for
functional inversion, and hence for a definition of the
``rotationally--symmetric non--holomorphic $N$--transform'' ---
even though for general non--Hermitian random matrices a
construction of a ``non--holomorphic $N$--transform'' remains
thus far unknown. This new $N$--transform is then conjectured to
be in a simple relation to the (usual) $N$--transform of the
Hermitian ensemble \smash{$\mathbf{X}^{\dagger} \mathbf{X}$}. In
a typical situation, the latter will be much more easily
solvable than the former, owing to the plethora of tools devised
in the Hermitian world, albeit the opposite may be true as well.
This is indeed the case here --- our derivation of
(\ref{eq:MPBasicEquation}), based on non--Hermitian planar
diagrammatics and Dyson--Schwinger's equations, is much more
involved than a simple application of the FRV multiplication
rule leading to (\ref{eq:MQBasicEquation}) --- and consequently,
the aforementioned hypothesis would provide a shortcut to avoid
complicated diagrammatics. To the best of our knowledge, this
would be the first use of Free Random Variables calculus to
compute the mean spectral density of a non--Hermitian product of
random matrices.

We have also suggested a heuristic model of the
finite--size behavior of the density of $\mathbf{P}$ near the
edge of the eigenvalues support (\ref{eq:EIG47}), deducing it
from analogous considerations~\cite{fh,k,ks} made for the Girko--Ginibre ensemble,
where this behavior is known analytically. It performs outstandingly well when checked against numerical simulations.

Let us also remark that one could
argue, as for square matrices, that the large--$N$ limit 
result is the same for elliptic
Gaussian ensembles~\cite{bjw}. We also believe that
one can further weaken the assumptions on the matrices involved, 
just requiring them to belong to the Gaussian universality class
of matrices having independent entries and fulfilling
the Pastur--Lindeberg condition~\cite{p} (the matrix analogue of the 
generalized central limit theorem in classical probability 
theory~\cite{LINDEBERG}). One unexpected
implication of such universality is that a product of random
matrices whose spectra do not necessarily display rotational
symmetry has an eigenvalue distribution 
which does possess rotational symmetry on the complex plane 
(\ie the average density depends only on $| \lambda |$).

Let us now list some possible applications of these
results to wireless telecommunication, quantum entanglement and
multivariate statistical analysis.

Information theory for wireless telecommunication has been
intensively developed in the past decade, after it had been
realized that in a number of situations the information transmission
rate can be increased by an introduction of multiple antenna channels,
known as the ``multiple--input, multiple--output'' (MIMO)
transmission links. The MIMO capacity for Gaussian channels has
been calculated in the pioneering work~\cite{t}, triggering
large activity in the field. Immediately, it became clear that
an appropriate language and methods to address this type of
problems are provided by random matrix theory (consult~\cite{tv}
for a review).
The model considered in our paper can be applied to a situation
of signals traveling over $L$ consecutive MIMO links. The signal
is first sent from \smash{$N_{1}$} transmitters via a MIMO link
to \smash{$N_{2}$} receivers, which then re--transmit it via a
new MIMO link to the subsequent \smash{$N_{3}$} receivers,
\emph{etc.} Clearly, the capacity will depend on these numbers
of intermediate re--transmitters; in particular, if any of the
\smash{$N_{l}$}'s is small, the capacity will be reduced. The
effective propagation is given by the matrix \smash{$\mathbf{P}
= \mathbf{A}_{L} \ldots \mathbf{A}_{2} \mathbf{A}_{1}$}. Such a model of multifold scattering per propagation path has been already proposed in \cite{MULLER}, where the moment generating function, the $M$--transform,
for ${\bf P}^{\dagger}\bf {P}$ was calculated.
Our result for the $M$--transform for ${\bf P}$ complements
this calculation.

Let us also mention that one could imagine a more
general situation, where MIMO links form a directed network ---
each directed link $l m$ representing a single MIMO channel
between \smash{$N_{l}$} transmitters and \smash{$N_{m}$}
receivers. (The previously discussed case
corresponds to a linear graph, $1 \to 2 \to \ldots \to L$.)
A complex directed network of MIMO links is somewhat similar to
the structures appearing in the context of quantum entanglement.
There, one considers graphs whose edges describe bi--partite
maximally entangled states, while vertices describe the couplings
between subsystems residing at the same vertex~\cite{cnz}. In
the simplest case of a graph consisting of a single link, it is
just a bi--partite entangled state. The corresponding density
matrix for a bi--partite subsystem is given by
\smash{$\mathbf{Q} = \mathbf{A}^{\dagger} \mathbf{A}$}, where
$\mathbf{A}$ is a rectangular matrix defining a pure state,
being a combination of the basis states in the subsystem,
\smash{$| \alpha_{a} \rangle$} and \smash{$| \beta_{b} \rangle$}
(see for instance~\cite{m}). One can easily find that linear
graphs with additional loops at the end vertices correspond. The
density matrix for the subsystem sitting in the end vertex is
given by \smash{$\mathbf{Q} = \mathbf{P}^{\dagger} \mathbf{P}$},
where \smash{$\mathbf{P} = \mathbf{A}_{1} \mathbf{A}_{2} \ldots
\mathbf{A}_{L}$}~\cite{cnz}. If all the subsystems are of the
same size, the average spectral distributions
\cite {bg,bbcc} of $\mathbf{Q}$ are known as the ``Fuss--Catalan
family''~\cite{a}; they can be obtained from
(\ref{eq:MQBasicEquation}) by setting all the \smash{$R_{l}$}'s
to $1$. However, if the subsystems have different sizes, one
needs to apply our general formula (\ref{eq:MQBasicEquation}).

Finally, another area of applications of our approach is related to
multivariate analysis. The main building block there is the Wishart ensemble, corresponding to $L=1$ in our formalism. The link between the spectral properties of ${\bf P}$ and $\bf {Q}$ may allow one to avoid the well--known bottleneck caused by the non--Hermiticity of time--lagging correlation functions. This issue will be discussed in a forthcoming publication.

\begin{acknowledgements}
We would like to thank R. A. Janik, B. Khoruzhenko,
and K. \.{Z}yczkowski for interesting discussions.
This work was partially supported by the Polish Ministry of
Science Grant No.~N~N202~229137~(2009--2012). AJ acknowledges
the support of Clico Ltd.
GL wishes to thank G. Montagna and O. Nicrosini for their kind support and
helpful suggestions; GL also acknowledges the Pavia University 
Ph.D. School in Physics for letting him be enrolled in the International
Ph.D. Programme.
\end{acknowledgements}

\end{document}